\documentclass[twocolumn,floats,showpacs,superscriptaddress,pre]{revtex4}
\usepackage{amssymb}
\usepackage{graphicx}
\usepackage{dcolumn}
\usepackage{amsmath}
\usepackage{bm}
\usepackage{epsfig}

\def\defi{{\buildrel \;def\; \over =}}
\newcommand{\be}{\begin{equation}}
\newcommand{\ee}{\end{equation}}

\newcommand{\mediaT}[1]{\left\langle #1 \right\rangle}

\newcommand{\media}[1]{\langle #1 \rangle}

\def\mod{ \mathop{\rm mod} }

\begin{document}

\title{Critical behavior and correlations on scale-free small-world networks.\\
Application to network design}
\author{M. Ostilli }
\affiliation{Departamento de F{\'\i}sica and I3N, Universidade de Aveiro, 3810-193 Aveiro,
Portugal}
\affiliation{Statistical Mechanics and Complexity Center (SMC), 
INFM-CNR SMC, Rome, Italy}
\author{A. L. Ferreira}
\affiliation{Departamento de F{\'\i}sica and I3N, Universidade de Aveiro, 3810-193 Aveiro,
Portugal}
\author{J. F. F. Mendes}
\affiliation{Departamento de F{\'\i}sica and I3N, Universidade de Aveiro, 3810-193 Aveiro,
Portugal}

\begin{abstract}
We analyze critical phenomena on networks generated as the union of hidden variables models 
(networks with any desired degree sequence)
with arbitrary  graphs. 
The resulting networks are general small-worlds similar to those \`a la Watts and Strogatz
but with a heterogeneous degree distribution.  
We prove that the critical behavior (thermal or percolative) remains completely 
unchanged by the presence of finite loops (or finite clustering).
Then, we show that,
in large but finite networks, 
correlations of two given spins 
may be strong, \textit{i.e.}, approximately power law like, at any temperature. 
Quite interestingly, if $\gamma$ is the exponent for the power law distribution of the vertex degree,
for $\gamma\leq 3$ and with or without short-range couplings, 
such strong correlations
persist even in the thermodynamic limit, contradicting the common
opinion that in mean-field models correlations always disappear
in this limit.
Finally,
we provide the optimal choice of rewiring
under which percolation phenomena in the rewired network are best performed;
a natural criterion to reach best communication features, at least in non congested regimes.
\end{abstract}

\maketitle

\email{ostilli@roma1.infn.it}


\section{Introduction} \label{intro} 
In the last decade, it has been recognized that
at the base of many complex systems, as diverse as those observed in nature, in technological, and in social sciences,
there is an ubiquitous presence of networks having certain universal topological 
features upon which the functionality of the system largely depends on \cite{DMAB}.
Essentially,
there are two basic topological features in these complex networks: 
scale-free and small-world. 
The former refers to the fact that  
the distribution of the links among the nodes is strongly heterogeneous,
in particular many networks have a power law distribution for the vertex degree, $\mathcal{P}(k)\sim k^{-\gamma}$;  
the latter refers to the fact that two randomly chosen nodes are at 
distance which, for $\gamma>3$, scales with the system size $N$ as slowly as $\log(N)/\log(b)$,
$b$ being the mean branching of the network while, for $\gamma\leq 3$ (where $b$ diverges), the average distance 
scales as $\log(\log(N))$ or even slower and the network is called ultra-small \cite{UltraSmall}. 
Network models have then been profusely studied over the years
and many fundamental results are by now well established and widespread \cite{Review,SND}.
The main assumption under these studies has been the tree-like hypothesis
thanks to which the generating function technique and the Bethe-Peierls (BP) method \cite{Mezard} can be applied
to get exact solutions for the percolative and thermal (at least in the ferromagnetic case) 
properties as well as the communication features of the system. 
However, the tree-like assumption is almost never satisfied in real-world networks.
For example, networks of friends, networks of neurons, the WWW, and the Internet, 
are just a few examples in which the average clustering coefficient $C$ \cite{DMAB,SND} is finite.
More precisely, whereas networks having a hierarchical structure share
a $k$ degree-dependent clustering coefficient of the form $C(k)\sim k^{-\alpha}$, with $\alpha \sim 1$, so that the
most connected (and most important) nodes are not clustered, there are other networks
having $C(k)\sim \mathop{O}(1)$ for almost any $k$ and for which clustering is important for all nodes. 
The former class includes \textit{e.g.} some social networks, language networks, 
the WWW, and the Internet at the autonomous system level, whereas the
latter class includes the Internet at the router level, the power grid, but also the brain.  
As discussed in \cite{Barabasi}, the reason for this difference is related to the fact that
in the second class wiring is expensive (economically or biologically) and the network, rather than 
hierarchically organized, is geographically organized. 
Finally, we recall that even in pure scale-free networks characterized by an exponent $\gamma\leq 3$,
the tree-like assumption is not true; such networks in fact contain many large cliques~\cite{BianconiMarsili}.

In the last few years there have been important progresses for the modeling 
of networks with loops \cite{HyperNewman,HyperBoguna,HyperBogunaII,HyperNewmanII,HyperGleeson}, however, 
such progresses were essentially confined to the cases 
in which the random graph can be seen as a tree-like hypergraph, or with a weak transitivity 
(\textit{i.e.}, with a small overlap between clusters).
It is then of fundamental importance to understand, in general, what is the role of the loops in 
complex networks from the point of view of collective behavior.
In the presence of loops how do the critical surface and the correlation functions change?  
Are the analytical results accumulated over ten years of research in complex networks 
robust with respect to the presence of loops? And if yes, to what extent?
We point out that the crucial question concerns the loops of finite length. 
In fact, in classical random graphs ($\gamma=\infty$) and in complex networks ($2<\gamma<\infty$) 
there are no finite loops 
(at least for $\gamma>3$; for a detailed discussion of the case $\gamma\leq 3$ see next Section)
since the length of the loops scales as $\log(N)$ and, 
as a consequence, one can say that the tree-like approximation in these models becomes exact for $N\to\infty$.
But in networks in which there are loops of finite length (for any $N$), 
due to the fact that the correlation length in all these models remains finite,
we are not allowed to neglect the short-loops neither near the critical point 
and, as a consequence, the (exact) solution to these issues is non trivial.

In this paper we address the above questions via the introduction of heterogeneous small-world networks,
a natural generalization of the ``classical'' small-world networks \cite{Watts}
which were introduced as intermediate systems lying between 
loopy-like (\textit{e.g.} finite-dimensional lattices) 
and tree-like networks.
In the classical ($\gamma=\infty$) small-world networks we have a homogeneous distribution of links among the sites,
but also a finite clustering coefficient, so that tree-like based techniques as the BP cannot be used
to solve, \textit{e.g.}, an Ising model defined on them.
By using a completely different approach it is however possible to solve exactly the
homogeneous small-world models at least in the paramagnetic phase (P) getting their exact critical
surface and behavior \cite{SW}. If $J_0$ is the coupling associated to a given graph $(\mathcal{L}_0,\Gamma_0)$,
which in particular may have short-range links and short loops of any kind,
and $J$ is the coupling associated to a number $Nc/2$ of additional 
uniformly spread long-range links (so that they alone would form
a classical random graph \cite{Classical}), 
then, for any $c>0$, the mean-field equation for these models is given by  
\begin{eqnarray}
\label{m0}
m=m_0(\beta J_0;c t m+\beta h), \quad t\defi \tanh(\beta J),
\end{eqnarray}
where $m_0(\beta J_0;\beta h)$ is defined as the average magnetization of the
model in the absence of the long-range connections, with a short-range coupling $J_0$ 
and in the presence of a generic external field $h$ at temperature $T=1/\beta$.
Eq.~(\ref{m0}) is a natural  generalization of the celebrated Curie-Weiss mean-field equation
$m=\tanh(\beta J m+\beta h)$, which is valid for $J_0=0$ and $c=N$. 
It is easy to check that, however, for any $J_0\geq 0$, the critical behavior of Eq.~(\ref{m0})
is classical, regardless of the local topology and clustering coefficient of the network (see also \cite{Hastings}).

In this paper we face the natural extension of Eq. (\ref{m0}) towards a large class of 
heterogeneous small-world networks generated by using hidden variables 
\cite{KimGoh0,CaldarelliHidden,BogunaHidden,NewmanHidden}, 
which have the little inconvenient that the resulting network has some small
degree-degree correlation \cite{SatorrasHidden}, but also the great advantage that the model is analytically solvable
even in the presence of loops (at least within our effective field theory).
After deriving the equation for the order parameter and the critical surface (thermal or bond-percolative),
we analyze the connected correlation functions in general and we show that
in these models, even for $J_0=0$, there are finite size corrections as strong as $1/N^\delta$,
with $\delta=(\gamma-2)/(\gamma-1)$ for $\gamma>3$, and $\delta=0$ for $\gamma\leq 3$,
contradicting the common opinion that in mean field models
the connected correlation functions always disappear in the thermodynamic limit.
Then we prove that the critical behavior (thermal or bond-percolative) 
on these networks is never affected by the presence of 
a local non tree-like structure, provided 
that the connectivity associated to such loopy structures is non heterogeneous.
This latter result has been already presented and discussed in a Letter \cite{SWSF} 
but limiting the proof to an infinitesimal coupling $J_0$, while
postponing to the present paper the general proof as well as the derivation
of the mean-field equation.
Nicely, meanwhile in \cite{BogunaC}, we find that  
a similar robustness theorem has been proved for both static
and growing networks embedded in a metric space when~$\gamma\leq 3$.

In tables \ref{t1} and \ref{t2} we summarize the state of the art 
reached about the analytical behavior of the Ising model built
on classical and complex random graphs. We stress that we mention
only the cases and the models where exact analytical calculations have been possible 
without any pretension to be exhaustive (in particular we do not mention
here the hierarchical models (random or not), where some exact analytical results are also possible).

\begin{table*}[chtbp]
\begin{tabular}{|l|l|l|l|}
\hline\hline
Ising on the ``Classical Random Graph'' Ref. \cite{Classical} (y. 1959) & $~~~~~~~\bar{m}$ & $~~\chi$ 
& Eq. for $t_c=\tanh(\beta_c J)$ \\ \hline\hline
$P(k)$ Poissonian (No Finite Loops; shortest loops scale as $~\mathop{O}(\log(N))$) & $\propto \tau^{1/2}$ & $\propto \tau^{-1}$ 
& $ct_c =1$\\ 
\hline\hline\hline
\hline\hline
Ising on the ``Configuration Network'' Refs. \cite{GZ} (y. 2002) & $~~~~~~~\bar{m}$ & $~~\chi$ 
& Eq. for $t_c=\tanh(\beta_c J)$ \\ \hline\hline
$\gamma >5,$ $\langle k^4\rangle_{_P} <\infty $ (shortest loops scale as $~\mathop{O}(\log(N))$) & $\propto \tau^{1/2}$ & & \\ \cline{1-2}
$\gamma =5,$ $\langle k^4 \rangle_{_P} =\infty ,$ $\langle k^2 \rangle_{_P} < \infty $ (shortest loops scale as $~\mathop{O}(\log(N))$)
& $\propto1/\ln \tau^{-1}$ & $\propto \tau^{-1}$ 
& $t_c=\frac{\media{k}_{_P}}{\media{k^2}_P-\media{k}_{_P}}$  \\ \cline{1-2}
$3<\gamma <5,$ $\langle k^4\rangle_{_P} =\infty ,$ $\langle k^2\rangle_{_P} <\infty $ (shortest loops scale as $~\mathop{O}(\log(N))$)
& $\propto \tau^{1/(\gamma -3)}$ & & \\ \hline
$\gamma =3,$ $\langle k^2\rangle_{_P} =\infty $ (Finite Loops)
& $\propto e^{-2T/\langle k\rangle_{_P}}$ &  
& $~~~~~~~~\beta_c\to 0$ \\ \cline{1-3} 
$2<\gamma <3,$ $\langle k^2 \rangle_{_P} =\infty $ (Finite Loops) & $\propto T^{-1/(3-\gamma )}$
& $\propto T^{-1}$ & \\
\hline\hline\hline
\hline\hline
Ising on the ``Static Network'' (Hidden Variables) Ref. \cite{KimStat} (y. 2005) & $~~~~~~~\bar{m}$ & $~~\chi$ 
& Eq. for $t_c=\tanh(\beta_c J)$ \\ \hline\hline
$\gamma >5,$ $\langle k^4\rangle_{_P} <\infty $ (shortest loops scale as $~\mathop{O}(\log(N))$) & $\propto \tau^{1/2}$ & & \\ \cline{1-2}
$\gamma =5,$ $\langle k^4 \rangle_{_P} =\infty ,$ $\langle k^2 \rangle_{_P} < \infty $ (shortest loops scale as $~\mathop{O}(\log(N))$)
& $\propto1/\ln \tau^{-1}$ & $\propto \tau^{-1}$ 
& $ct_cN \sum_{i}p_i^2=1$  \\ \cline{1-2}
$3<\gamma <5,$ $\langle k^4\rangle_{_P} =\infty ,$ $\langle k^2\rangle_{_P} <\infty $ (shortest loops scale as $~\mathop{O}(\log(N))$)
& $\propto \tau^{1/(\gamma -3)}$ & & \\ \hline
$\gamma =3,$ $\langle k^2\rangle_{_P} =\infty $ (Finite Loops)
& $\propto e^{-2T/\langle k\rangle_{_P}}$ &  
& $~~~~~~~~\beta_c\to 0$ \\ \cline{1-3} 
$2<\gamma <3,$ $\langle k^2 \rangle_{_P} =\infty $ (Finite Loops) & $\propto T^{-1/(3-\gamma )}$ 
& $\propto T^{-1}$ & \\
\hline\hline\hline
\hline\hline
Ising on the ``Classical SW Networks'' Refs. \cite{SW,Hastings} (y. 2003,08) & $~~~~~~~\bar{m}$ & $~~\chi$ 
& Eq. for $t_c=\tanh(\beta_c J)$ \\ \hline\hline
$P(k)$ Poissonian + additional arbitrary links (Loops of any length) & $\propto \tau^{1/2}$ & $\propto \tau^{-1}$ 
& $ct_c \tilde{\chi}_{0}
\left(\beta_c J_0;0\right)=1$ \\ 
\hline\hline\hline
\hline\hline
Ising on the ``Heterogeneous SW N.'' Ref. \cite{SWSF} (y. 2010) and Present Work & $~~~~~~~\bar{m}$ & $~~\chi$ 
& Eq. for $t_c=\tanh(\beta_c J)$ \\ \hline\hline
$\gamma >5,$ $\langle k^4\rangle_{_P} <\infty $ (Arbitrary Loops of any length) & $\propto \tau^{1/2}$ & & \\ \cline{1-2}
$\gamma =5,$ $\langle k^4 \rangle_{_P} =\infty ,$ $\langle k^2 \rangle_{_P} < \infty $ (Arbitrary Loops of any length)
& $\propto1/\ln \tau^{-1}$ & $\propto \tau^{-1}$ 
& $ct_cN \sum_{i,j}\tilde{\chi}_{0;i,j}p_ip_j=1$ \\ \cline{1-2}
$3<\gamma <5,$ $\langle k^4\rangle_{_P} =\infty ,$ $\langle k^2\rangle_{_P} <\infty $ (Arbitrary Loops of any length)
& $\propto \tau^{1/(\gamma -3)}$ & & \\ \hline
$\gamma =3,$ $\langle k^2\rangle_{_P} =\infty $ (Arbitrary Loops of any length)
& $\propto e^{-2T/\langle k\rangle_{_P}}$ &  
& $~~~~~~~~\beta_c\to 0$ \\ \cline{1-3} 
$2<\gamma <3,$ $\langle k^2 \rangle_{_P} =\infty $ (Arbitrary Loops of any length) & $\propto T^{-1/(3-\gamma )}$ 
& $\propto T^{-1}$ & \\
\hline\hline\hline
\hline\hline
Ising on ``Static and Growing Spatial Networks'' Ref. \cite{BogunaC} (y. 2011) & $~~~~~~~\bar{m}$ & $~~\chi$ 
& Eq. for $t_c=\tanh(\beta_c J)$ \\ \hline\hline
Self Similarity $2<\gamma <3,$ $\langle k^2 \rangle_{_P} =\infty $ (Loops in a Metric Space) 
& $\propto T^{-1/(3-\gamma )}$ & $\propto T^{-1}$ & $\beta_c\to 0$ \\
\hline\hline
\end{tabular}
\caption{ Critical behavior of the unweighted magnetization 
$\overline{m}=\sum_i\overline{\mediaT{\sigma_i}}/N$ ($\tau \equiv 1-T/T_c$), 
the susceptibility $\chi$, and the equation for the critical temperature
of the Ising model built on several Network Models: the Classical Random Graph;
the Configuration Model (\textit{i.e.}, the maximally random graph under the constraint that the degree distribution 
is a given one, $P(k)\sim k^{-\gamma}$); 
the Static Model (a Hidden Variables Model with weights $p_i\propto i^{-\mu}$, where 
$\mu\in[0,1)$ is such that $P(k)\sim k^{-\gamma}$); 
the Classical Small World Network built by overlapping the Classical Random Graph with additional links
associated to an arbitrary graph $(\mathcal{L}_0,\Gamma_0)$;   
the Heterogeneous Small World Networks (\cite{SWSF} ($J_0$ infinitesimal) and present work ($J_0$ arbitrary)) 
built by overlapping Hidden Variables Models 
with additional links associated to an arbitrary graph $(\mathcal{L}_0,\Gamma_0)$;   
the Spatial Network Model for $\gamma<3$ (where self-similarity applies). The data in parenthesis stands
for the year of publication. We use the notation $c=\media{k}_{_P}$.} 
\label{t1}
\end{table*}

\begin{table*}[chtbp]
\begin{tabular}{|l|l|}
\hline\hline
Ising on the ``Configuration Network'' Ref. \cite{Corr} (y. 2005) & Correlations $\tilde{\chi}_l$ of two spins 
at distance $l$; $N\to\infty$ \\ \hline\hline
$\gamma >5,$ $\langle k^4\rangle_{_P} <\infty $ (length of shortest loops scale as $~\mathop{O}(\log(N))$) & 
$\tilde{\chi}_l\sim t^l$, ~ ($t=\tanh(\beta J)$) \\ \hline
$\gamma =5,$ $\langle k^4 \rangle_{_P} =\infty ,$ $\langle k^2 \rangle_{_P} < \infty $ (shortest loops scale as $~\mathop{O}(\log(N))$) & $\tilde{\chi}_l\sim t^l$ \\ \hline
$3<\gamma <5,$ $\langle k^4\rangle_{_P} =\infty ,$ $\langle k^2\rangle_{_P} <\infty $ (shortest loops scale as $~\mathop{O}(\log(N))$) & $\tilde{\chi}_l\sim t^l$ \\ \hline
$\gamma =3,$ $\langle k^2\rangle_{_P} =\infty $ (Finite Loops) 
& $\tilde{\chi}_l $ ? \\ \hline
$2<\gamma <3,$ $\langle k^2 \rangle_{_P} =\infty $ (Finite Loops)  & $\tilde{\chi}_l $ ? \\
\hline\hline\hline
\hline\hline
Ising on the ``Classical SW Networks'' Refs. \cite{SW} (y. 2008) 
& Correlations $\tilde{\chi}_{ij}$ of two given spins $i$ and $j$; $N$ finite \\ \hline\hline
$P(k)$ Poissonian + additional arbitrary links (Arbitrary Loops) & 
$\tilde{\chi}_{ij}=\tilde{\chi}_{0;i,j}+ 
\mathop{O}\left(\frac{ct}{N}
\frac{\left[\tilde{\chi}_{0}\right]^2}
{1-ct\tilde{\chi}_{0}}\right)$ \\
\hline\hline\hline
\hline\hline
Ising on the ``Heterogeneous SW N.'' (Present Work) 
& Correlations $\tilde{\chi}_{ij}$ of two given spins $i$ and $j$; $N$ finite\\ \hline\hline
$\gamma >3,$ $\langle k^4\rangle_{_P} <\infty $ (Arbitrary Loops of any length) & 
$\tilde{\chi}_{ij}=\tilde{\chi}_{0;i,j}+ 
\mathop{O}\left(\frac{t}{t_c}\frac{1}{1-t/t_c}\frac{(ij)^{-1/(\gamma-1)}}{N^{(\gamma-3)/(\gamma-1)}}\right)$ \\
\hline
$3\geq \gamma >2,$ $\langle k^4\rangle_{_P} =\infty $ (Arbitrary Loops of any length) & 
$\tilde{\chi}_{ij}=\tilde{\chi}_{0;i,j}+ 
\mathop{O}\left((ij)^{-1/(\gamma-1)}\right)$ \\
\hline\hline
\end{tabular}
\caption{ Connected Correlation functions $\tilde{\chi_l}$ 
(connected correlation of two randomly chosen spins at given distance $l$) 
and $\tilde{\chi}_{ij}$ (average connected correlation of two given spins $i$ and $j$, see Sec. IIC)
of the Ising model built on: 
the Configuration Model; 
the Classical Small World Network built by overlapping the Classical Random Graph with additional links
associated to an arbitrary graph $(\mathcal{L}_0,\Gamma_0)$
($\tilde{\chi}_{0;i,j}$ stands for the connected correlation function associated to $(\mathcal{L}_0,\Gamma_0)$);   
the Heterogeneous Small World Networks (present work) built by overlapping Hidden Variables Models 
with $(\mathcal{L}_0,\Gamma_0)$. In the Heterogeneous Small World Networks 
the formula for $\tilde{\chi}_{ij}$ is valid when  
the underlying network $(\mathcal{L}_0,\Gamma_0)$ has dimension $d_0\leq 1$. For the general case
the dependence on $i$ and $j$ is more complicated (see Eq. (\ref{THEOC2b})), but the dependence on $N$ is the same.
For what concerns the configuration network model, 
for $\gamma\leq 3$, in Ref. \cite{Corr} it is speculated that $\tilde{\chi}_l\sim 0$
for $l$ not too small, while for $l\sim \mathop{O}(1)$ one has $\tilde{\chi}_l\sim t^l$; a result which is not in contradiction with our
achievement (see Sec. IIC). However, in Ref. \cite{Corr} $\tilde{\chi}_l$ was obtained directly by using the tree-like
assumption which, for $\gamma\leq 3$, is wrong even in the configuration model (in fact the formulas of Ref. \cite{Corr}
give $\tilde{\chi}_l=0$ for any $l\geq 2$, which is clearly not exact). 
See Ref. \cite{BianconiMarsili} and discussion [44]. Apart from technical details, 
the main point we stress is that in previous works was not possible to see that: 
\textit{i)} finite size effects are always strong in scale-free networks;  
\textit{ii)} for $\gamma>3$ they decay as slowly as $\mathop{O}(1/N^{(\gamma-3)/(\gamma-1)})$;  
\textit{iii)} finite size effects persist even in the thermodynamic limit when $\gamma \leq 3$ and,
as a consequence, correlations of two given spins can be strong (power law like) when $\gamma \leq 3$.}
\label{t2}
\end{table*}

Finally, as an application, given the desired degree sequence and the graph $(\mathcal{L}_0,\Gamma_0)$, 
we find that the equation for the critical surface leads
to an optimization problem consisting in finding
the rewiring of the additional links that provides the minimal percolating point;
a criterion which amounts to finding the rewiring that provides the 
best communication performance at the minimal cost in the absence of congestion.
This optimization problem in general is an NP-hard problem, however, we provide heuristic solutions whose
effectiveness depends on how much the network is structured in communities (if any),
and we show that the use of the formula for the critical surface
is always exponentially (in $N$) convenient with respect to a direct inspection
of the network, even in the worst case scenario in which there is no
community structure at all.
%

%
%

\section{Random Ising models built on heterogeneous small-world networks}
\label{models}
\subsection{The model}
The family of models we shall consider are built as follows.
Let $(\mathcal{L}_0,\Gamma_0)$ be any graph, 
$\mathcal{L}_0$ and $\Gamma_0$ being the set of vertices $i=1,\ldots,N$ and links $(i,j), ~i<j$, respectively.
Let us consider the Ising model defined on the graph $(\mathcal{L}_0,\Gamma_0)$ 
with a fixed coupling $J_0$ and in the presence of an arbitrary external field $\{h_i\}$ 
\begin{eqnarray}
\label{H0}
H_0= -J_{0}\sum_{(i,j)\in \Gamma_0}\sigma_{i}\sigma_{j}-\sum_i h_i\sigma_i.
\end{eqnarray}
We will call this \textit{the pure model}. 
Let us now consider the model obtained by removing randomly some links of the graph $(\mathcal{L}_0,\Gamma_0)$
and by adding new links as follows. Let us indicate with $c_{0;i,j}=0,1$ the adjacency matrix 
of the new graph in which some links of $\Gamma_0$ have been removed.
Given an ensemble $\mathcal{C}$ 
of random graphs $\bm{c}$, $\bm{c}\in\mathcal{C}$,
whose links are determined by the adjacency matrix elements $c_{i,j}=0,1$,
we define our \textit{heterogeneous small-world model}, through the following Hamiltonian 
\begin{eqnarray}
\label{H}
H_{\bm{c}_0,\bm{c},\bm{J}_0,\bm{J}}&\defi& -\sum_{(i,j)\in \Gamma_0}c_{0;i,j}J_{0;i,j}\sigma_{i}\sigma_{j}
 -h\sum_i \sigma_i\nonumber \\
&& -\sum_{i<j} c_{ij}{J}_{ij}\sigma_{i}\sigma_{j}.
\end{eqnarray}
The variables
$c_{i,j}$ specify whether a ``long-range'' link between the sites
$i$ and $j$ is present ($c_{i,j}=1$) or absent ($c_{i,j}=0$), whereas
the variables $c_{0;i,j}$ specify whether a link $(i,j)\in\Gamma_0$
has been removed ($c_{0;i,j}=0$) or not ($c_{0;i,j}=1$).
The $J_{i,j}$'s are the random couplings of the given link $(i,j)$ and similarly
for the $J_{0;i,j}$'s for the links of $\Gamma_0$.
All the above random variables are assumed to be independent. 
For the $J_{0;i,j}$'s and the $J_{i,j}$'s we will not assume any particular distribution, 
while for the $c_{0;i,j}$'s and the $c_{i,j}$'s we assume respectively the following probabilities 
\begin{eqnarray}
\label{p0}
p_0(c_{0;i,j})=(1-p)\delta_{c_{0;i,j},1}+p\delta_{c_{0;i,j},0},
\end{eqnarray}
\begin{eqnarray}
\label{hidden}
 p_{ij}(c_{ij})=
f\left(p_i,p_j\right)\delta_{c_{ij},1}+
(1-f\left(p_i,p_j\right))\delta_{c_{ij},0},
\end{eqnarray}
where $p\in[0,1]$, and the $\{p_i\}$ are a set of hidden variables
\cite{KimGoh0,CaldarelliHidden,BogunaHidden,NewmanHidden}~\footnote{Usually
the hidden variables are represented not with the $\{p_i\}$, but with the set
$\{\theta_i\}$ where $\theta_i=\sqrt{cN}p_i$.} each proportional
to the average degrees $\{\bar{k}_i\}$ of the graph $\bm{c}$ 
of the nodes $i=1,\ldots,N$ (\textit{i.e.}, the degrees in the absence of the graph 
$(\mathcal{L}_0,\Gamma_0)$).
Usually the hidden variables depend on one (or more) continuous parameters $\mu\in \mathcal{I}$, and on $N$.
Given the mean degree $c>0$ (so that in average there are in total $cN/2$ bonds) of the graph $\bm{c}$,
we will assume that for a continuous subset $\mathcal{J}\subset\mathcal{I}$, asymptotically in $N$
we can write 
\begin{eqnarray}
\label{hidden1}
f\left(p_i,p_j\right)=cN p_ip_j,
\end{eqnarray}
where
\begin{eqnarray}
\label{hidden2}
c\defi\sum_i \frac{\bar{k}_i}{N}.
\end{eqnarray}
For the validity of the results we present in the next Section we require 
the number of links $(i,j)$ for which Eq. (\ref{hidden1}) is not true, to be less than $\mathop{O}(N^\alpha)$,
as long as $\alpha<1$. 
We will prove in fact that, in the thermodynamic limit,  
the free energy of the model (see below) is not affected by the presence
of the $\mathop{O}(N^\alpha)$ links for which Eq. (\ref{hidden1}) is not true if $\alpha<1$. 
As a probability, Eq. (\ref{hidden1}) for $f(p_i,p_j)$ will be
manifestly violated in $\mathcal{I}\setminus\mathcal{J}$ whenever $cNp_ip_j>1$.
Note that, if $p_i\neq 0$ for any given $N$ (a requirement which is true for any graph
in which there are not isolated nodes), for $N\to\infty$ the terms $cN p_ip_j$
tend either to $0$ or to $\infty$, therefore, the number of links $(i,j)$ for which
Eq. (\ref{hidden1}) is not true for $N$ large approaches 
\begin{eqnarray}
\label{hidden3}
\mathcal{N}_N\defi \sum_{i<j}\theta\left(cN p_ip_j-1\right),
\end{eqnarray}
where $\theta(x)=0$ or 1 if $x<0$ or $x\geq 0$, respectively.
In Appendix A we show that,
if in the ensemble $\mathcal{C}$ the probability $p(k)$ to have a vertex with degree $k$ scales, for $k$ large, 
as a power law $p(k)\sim k^{-\gamma}$, then:
\begin{eqnarray}
\label{hidden3cbis}
\mathcal{N}_N <\frac{N^{2-\gamma}c^{1-\gamma}}{2(\gamma-1)}\log(N),
\end{eqnarray}
so that the requirement $\mathcal{N}_N=\mathop{O}(N^\alpha)$ with $\alpha<1$ 
is equivalent to have $\gamma>2$, and, for $N$ large but finite, the error we make per spin in neglecting
these $\mathop{O}(N^\alpha)$ contributions is $\mathop{O}(N^{1-\gamma}\log(N))$ for $\gamma>2$~
\footnote{  
It should be noted that, as a matter of fact, from the analysis performed on $\mathcal{J}$
one is allowed to make the analytic continuation to get the results in the full set $\mathcal{I}$.
Notice the strict analogy with what is usually (tacitly) done in the
Ising model defined on the configuration model \cite{GZ}: 
one uses the local tree-like ansatz to derive the equation for the order
parameter in the region $\gamma>3$, then one extrapolates by analytic continuation the result
to the region $3\geq\gamma>2$ where the tree ansatz is wrong even locally. 
The reason why the analytic continuation works is the same as ours: the extensive free energy
does not depend on the number of contributions for which the tree-like ansatz is wrong since this
number grows less slowly than $\mathop{O}(N^\alpha)$ with some $\alpha<1$. 
It should be however recalled that the tree-like
ansatz used to get directly local quantities loop sensitive, as the spin-spin correlations, would
lead to a completely wrong result for $3\geq\gamma>2$. The proper way to get the
spin-spin correlation consists in solving the model for $\gamma>3$ in the presence
of a non uniform external field and then to analytically continue the result to the range $3\geq \gamma>2$.}.
As an example of a scale-free model we can consider the choice   
\begin{eqnarray}
\label{stat}
f\left(p_i,p_j\right)=1-e^{-cNp_ip_j}
\end{eqnarray}
\begin{eqnarray}
\label{stat1}
 p_i\defi \frac{i^{-\mu}}{\sum_{j\in \mathcal{L}_0}j^{-\mu}}\simeq
\frac{i^{-\mu}(1-\mu)}{N^{1-\mu}},
\end{eqnarray}
where $\mu\in[0,1)$.
Equations (\ref{stat})-(\ref{stat1}) define the static model introduced in \cite{KimGoh0}.
Note that the so called fermionic constraint that avoids to have multiple bonds,
is automatically satisfied by Eq. (\ref{hidden}). As has been shown,
this constraint leads to some weak dis-assortative degree-degree correlations for $\mu>1/2$ \cite{SatorrasHidden}.
In the thermodynamic limit $N\to\infty$, for $\mu\in (0,1)$, Eqs. (\ref{stat})-(\ref{stat1}) lead to a
number of long range connections per site distributed according
to a power law with mean $c$ and exponent $\gamma$ given by 
\begin{eqnarray}
\label{stat2}
\gamma=1+\frac{1}{\mu},
\end{eqnarray}
so that $\gamma\in(2,\infty)$. 
For $\mu\in(0,1/2)$ ($\gamma>3$) Eq. (\ref{stat}) takes the simpler form (\ref{hidden1}) 
while for $\mu\in[1/2,1)$ ($2<\gamma\leq 3$) Eq. (\ref{stat}) can be written as Eq. (\ref{hidden1})
only when $i$ and $j$ are sufficiently distant, $ij\gg N^{2-1/\mu}$, while for lower distances,
$ij\ll N^{2-1/\mu}$, we have $p_{ij}(c_{ij}=1)\simeq 1$. 
 
The free energy $F$ and the averages $\overline{\media{\mathcal{O}}^l}$, with $l=1,2$,
are defined in the usual (quenched) way as ($\beta=1/T$)
\begin{eqnarray}
\label{logZ}
-\beta F\defi \sum_{\bm{c}_0,\bm{c}} 
P(\bm{c}_0,\bm{c})\int d\mathcal{P}\left(\bm{J}_0,\bm{J}\right)
\log\left(Z_{\bm{c}_0,\bm{c},\bm{J}}\right)
\end{eqnarray} 
and 
\begin{eqnarray}
\label{O}
\overline{\media{\mathcal{O}}^l}\defi 
\sum_{\bm{c}_0,\bm{c}} P(\bm{c}_0,\bm{c}) \int d\mathcal{P}\left(\bm{J}_0,\bm{J}\right)
\media{\mathcal{O}}_{\bm{c}_0,\bm{c},\bm{J}_0,\bm{J}}^l, \quad l=1,2
\end{eqnarray} 
where $Z_{\bm{c}_0,\bm{c},\bm{J}_0,\bm{J}}$ 
is the partition function of the quenched system
\begin{eqnarray}
\label{Z}
Z_{\bm{c}_0,\bm{c},\bm{J}_0,\bm{J}}= \sum_{\{\sigma_{i}\}}
e^{-\beta H_{\bm{c}_0,\bm{c},\bm{J}_0,\bm{J}}\left(\{\sigma_i\}\right)}, 
\end{eqnarray} 
$\media{\mathcal{O}}_{\bm{c}_0,\bm{c},\bm{J}_0,\bm{J}}$ the Boltzmann-average 
of the quenched system ($\media{\mathcal{O}}$ depends on the
given realization of $\bm{J}$, $\bm{J}_0$, $\{\bm{c}_0\}$ and $\bm{c}$:
$\media{\mathcal{O}}=\media{\mathcal{O}}_{\bm{c}_0,\bm{c};\bm{J}_0,\bm{J}}$;
for shortness we later will omit to write these dependencies)
\begin{eqnarray}
\label{OO}
\media{\mathcal{O}}_{\bm{c}_0,\bm{c},\bm{J}_0,\bm{J}}\defi \frac{\sum_{\{\sigma_i\}}\mathcal{O}e^{-\beta 
H_{\bm{c}_0,\bm{c},\bm{J}_0,\bm{J}}\left(\{\sigma_i\}\right)}}{Z_{\bm{c}_0,\bm{c},\bm{J}_0,\bm{J}}}, 
\end{eqnarray} 
and $d\mathcal{P}\left(\bm{J}_0,\bm{J}\right)$ and $P(\bm{c}_0,\bm{c})$ 
are product measures 
given in terms of arbitrary measures (all normalized to 1) 
for the short- and long-range couplings 
$d\mu_0(J_{0;i,j})\geq 0$, $d\mu(J_{i,j})\geq 0$, and in
terms of the introduced link probabilities (Eqs. (\ref{p0}) and (\ref{hidden}))
$p_0(c_{i,j})\geq 0$, and $p_{ij}(c_{i,j})\geq 0$: 
\begin{eqnarray}
\label{dP}
d\mathcal{P}\left(\bm{J}_0,\bm{J}\right)\defi \prod_{(i,j),i<j} 
d\mu\left( {J}_{i,j} \right)\prod_{(i,j)\in\Gamma_0}d\mu_0\left( {J}_{0;i,j} \right),
\end{eqnarray}
\begin{eqnarray}
\label{Pg}
P(\bm{c}_0,\bm{c})\defi \prod_{(i,j),i<j} p_{ij}(c_{i,j})\prod_{(i,j)\in\Gamma_0}p_0(c_{0;i,j}).
\end{eqnarray}

\subsection{A note on small-world networks \`a la Watts and Strogatz}
The class of our small-world scale-free models given by Eqs. (\ref{H0})-(\ref{hidden2}) is very general.
Note in particular that the graph $(\mathcal{L}_0,\Gamma_0)$ is completely arbitrary 
and can contain closed paths of any length. We stress that the resulting
network, union of the graph $(\mathcal{L}_0,\Gamma_0)$ in which each link is removed
with a probability $p$, with the scale-free random graph $\bm{c}$, can be seen as a scale-free 
gran canonical generalization
of the original small-world graph of Watts and Strogatz \cite{Watts},
though we here do not perform a true rewiring.
Since we let the probability $p\in[0,1]$ and the mean $c\in(0,\infty)$ arbitrary, our way
to build small-world networks is more general even for the non scale-free case $\mu=0$ ($\gamma=\infty$).
However, we can always restrict our class of small-world networks to the ones having a total
average connectivity which does not change with $p$
by choosing $c$ such that the total number of links of the graph $\bm{c}$
is equal to the total number of removed links of $(\mathcal{L}_0,\Gamma_0)$.
Up to corrections $\mathop{O}(1/\sqrt{N})$ we can accomplish this for any sample
by simply taking $c=c_0 p$, where $c_0$ is the average connectivity of $(\mathcal{L}_0,\Gamma_0)$.
We anticipate however that the critical behavior of these models is not affected
by any particular choice of $p$ and $c$, the only condition being $c>0$.  
In fact, as we will see soon in Sec. III, the sole role of the parameter $p$
is to give a renormalized effective coupling $J_0(p)$ to be used as though
we had the original graph $(\mathcal{L}_0,\Gamma_0)$ with no removed links.
Since the class of universality does not depend on $J_0$ as soon as $c>0$ it follows
that the critical behavior of this class of generalized small-world models
is the same for any $p\in[0,1]$ as soon as $c>0$~
\footnote{In the non scale-free case, 
the similarity between small-world models obtained by pure rewiring or pure addition of links
has been already speculated by several authors but never proved.}.

\subsection{Correlations of two given spins and correlations of two spins at given chemical distance}
The quantities of major interest are the averages, and the quadratic
averages, of the correlation functions 
which for shortness will be indicated by ${{C}}^{(\mathrm{1})}$ and
${{C}}^{(\mathrm{2})}$. For example, the following are non connected correlation functions
of order $k$:
\begin{eqnarray}
\label{CF}
{{C}}^{(\mathrm{1})}&=&\overline{\media {\sigma_{i_1}\ldots \sigma_{i_k}} }, \\
\label{CG}
{{C}}^{(\mathrm{2})}&=&\overline{\media {\sigma_{i_1}\ldots \sigma_{i_k}}^2 },
\end{eqnarray} 
where $k\geq 1$ and the indices $i_1,\ldots,i_k$ are supposed all different. 
For shortness we will keep using the symbols ${{C}}^{(\mathrm{1})}$ 
and ${{C}}^{(\mathrm{2})}$
also for the connected correlation function since
they obey to the same rules of transformations.
We point out that the set of indices $i_1,\ldots,i_k$ is
fixed along the process of the two averages, with respect to the couplings (\ref{dP}) 
and to the graph realizations (\ref{Pg}). 
This implies in particular
that, given the spin with index $i$ and the spin with index $j$, once the
averages have been performed,
their chemical distance remains undefined, while 
the only meaningful distance between $i$ and $j$, is the distance  
defined over $\mathcal{L}_0$, which we will indicate as $||i-j||_{_{_0}}$.
Interestingly, for the cases in which $(\mathcal{L}_0,\Gamma_0)$ is a regular 
lattice, $||i-j||_{_{_0}}$ is an Euclidean distance.  
Therefore, throughout this paper, it must be kept in mind that, 
for example, ${{C}}^{(\mathrm{1})}(||i-j||_{_{_0}})=
\overline{\media{\sigma_{i}\sigma_{j}} }$ 
is very different from the correlation function ${{G}}^{(\mathrm{1})}(l)$ 
of two points at a fixed chemical
distance $l$, \textit{i.e.}, the minimum number 
of links to join two points
among both the links of $\Gamma_0$ and the links of the random graph realization $\bm{c}$.
In fact, if, \textit{e.g.} for the homogeneous case $p_i\equiv 1/N$ with $J_0=0$, one considers all the possible 
realizations of the Poisson graph, and then all the possible
distances $l$ between two given points $i$ and $j$, one has   
\begin{eqnarray}
\label{CF1}
{{C}}^{(\mathrm{1})}(||i-j||_{_{_0}})&=&
\overline{\media {\sigma_{i}\sigma_{j}} }-\overline{\media
  {\sigma_{i}}} \overline{\media {\sigma_{j}} }
\nonumber \\
&=&\sum_{l=1}^N P_N(l){{G}}^{(\mathrm{1})}(l)
\end{eqnarray} 
where here $P_N(l)$ is the probability that, in the system with $N$ spins, 
the shortest path between the vertices $i$ and $j$ has length $l$.
If we now use ${{G}}^{(\mathrm{1})}(l)$
$\sim (\tanh(\beta J))^l$ \cite{Corr} (in the P region holds the exact equality) and the fact that
the average of $l$ with respect to $P_N(l)$ 
is of the order $\log(N)$, we see that the two point connected
correlation function (\ref{CF1}) goes to 0 in the thermodynamic limit. 
Similarly, in the Poissonian graph, all the connected
correlation functions defined in this way are zero in the thermodynamic limit.
However, as we will see in Sec. IIID, this independence of the variables holds only if $J_0=0$ and $\gamma>3$.
Furthermore, even for $\gamma>3$ finite size effects may result in strong correlations in the finite network.

\section{An effective field theory}
\subsection{The self-consistent equation}
Depending on the temperature T, and 
on the parameters $\mu$ and those of the probability distributions, $d\mu(\cdot)$ and $d\mu_0(\cdot)$,
the small-world model may stably stay either in the paramagnetic (P), in the ferromagnetic (F), 
or in the spin-glass (SG) phase.
In our approach for the F and SG phases there are two natural order parameters
that will be indicated by $m^{(\mathrm{F})}$ and $m^{(\mathrm{SG})}$.
Similarly, for any correlation function, quadratic or not, there are
two natural quantities 
indicated by $C^{(\mathrm{F})}$ and $C^{(\mathrm{SG})}$, and that in turn
will be calculated in terms of $m^{(\mathrm{F})}$ and $m^{(\mathrm{SG})}$,
respectively. 
To avoid confusion, it should be kept in mind that
in our approach,
for any observable $\mathcal{O}$,
there are - in principle - always 
two solutions that we label as F and SG,
but, as we shall discuss soon, for any temperature,
only one of the two solutions is stable and useful 
in the thermodynamic limit.

In the following, we will use the label $\mathop{}_0$ 
to specify that we are referring
to the pure model with Hamiltonian (\ref{H0}).
Note that all the equations presented in this paper 
have meaning and usefulness also for sufficiently large but
finite size $N$. For shortness we shall often omit to write the dependence on $N$.
 
Let $m_{0i}(\beta J_0,\{\beta h_j\})$ be the stable magnetization 
of the spin $i$ in the pure model (\ref{H0}) with coupling $J_0$ and in the presence of a
generic external field $\{h_j\}$ at inverse temperature $\beta$. 
In Appendix B we prove that 
the order parameter 
$m^{(\mathrm{F})}$ or $m^{(\mathrm{SG})}$ of the model defined in Eqs. (\ref{H})-(\ref{hidden2}),
with the condition $\alpha<1$ (equivalent to $\gamma>2$)
satisfies the following self-consistent equation
\begin{eqnarray}
\label{THEOa}
m^{(\Sigma)}=\sum_i m_{0i}(\beta J_0^{(\Sigma)};
\{Np_jct^{(\Sigma)}m^{(\Sigma)}+\beta h\})p_i,
\end{eqnarray} 
where the effective fields $t^{(\mathrm{F})}$, $t^{(\mathrm{SG})}$,
and couplings $J_0^{(\mathrm{F})}$ and $J_0^{(\mathrm{SG})}$, are given by
\begin{eqnarray}
\label{THEOb}
t^{(\mathrm{F})}= \int d\mu(J)\tanh(\beta J),
\end{eqnarray} 
\begin{eqnarray}
\label{THEOc}
t^{(\mathrm{SG})}= \int d\mu(J)\tanh^2(\beta J),
\end{eqnarray} 
\begin{eqnarray}
\label{THEOd}
\tanh(\beta J_0^{(\mathrm{F})})= (1-p)\int d\mu_0(J_0)\tanh(\beta J_0),
\end{eqnarray} 
and
\begin{eqnarray}
\label{THEOe}
\tanh(\beta J_0^{(\mathrm{SG})})= (1-p)\int d\mu_0(J_0)\tanh^2(\beta J_0).
\end{eqnarray}
Note that $|J_0^{(\mathrm{F})}|>J_0^{(\mathrm{SG})}$.
For later use we introduce also the short notations
\begin{eqnarray}
\label{THEOdS}
t_0^{(\mathrm{F})}\defi \tanh(\beta J_0^{(\mathrm{F})}),
\end{eqnarray} 
and
\begin{eqnarray}
\label{THEOeS}
t_0^{(\mathrm{SG})}\defi \tanh(\beta J_0^{(\mathrm{SG})}).
\end{eqnarray}
The meaning of the order parameters $m^{(\Sigma)}$ is quite natural being given by
\begin{eqnarray}
\label{orderpar}
\left(m^{(\Sigma)}\right)^{l_\Sigma}=\sum_i p_i \overline{\media{\sigma_{i}}^{l_\Sigma}},
\end{eqnarray} 
where $l_\Sigma=1,2$ for $\Sigma=$F or SG, respectively.
%
%

The free energy density $f^{(\Sigma)}$ coming from Eq. (\ref{logZ}) involves a generalized 
Landau free energy density $L^{(\Sigma)}$ from
which it differs only for trivial terms independent from $m^{(\Sigma)}$.
The complete expression for $f^{(\Sigma)}$ in terms of $L^{(\Sigma)}$ is reported in Appendix C.
The term $L^{(\Sigma)}$ reads ($\beta f^{(\Sigma)}=$ trivial terms $+L^{(\Sigma)}/l^{(\Sigma)}$, with $l^{(\Sigma)}=1,2$
for $\Sigma=$F, SG, respectively), and is given by
\begin{eqnarray}
\label{THEOll}
&& L^{(\Sigma)}(m^{(\Sigma)})\defi
\frac{ct^{(\Sigma)}\left(m^{(\Sigma)}\right)^2}{2}+\nonumber \\
&& \beta f_0\left(\beta J_0^{(\Sigma)},\{Np_jct^{(\Sigma)}m^{(\Sigma)}+\beta h\}\right),
\end{eqnarray} 
$f_0(\beta J_0,\{\beta h_i\})$ being the free energy density of the pure model (\ref{H0}). 
For given $\beta$, among all the possible solutions of Eqs. (\ref{THEOa}), in the thermodynamic
limit, for both $\Sigma$=F and $\Sigma$=SG, 
the true solution $\bar{m}^{(\Sigma)}$, or leading solution, 
is the one that minimizes $L^{(\Sigma)}$.

Finally, let $k$ be the order of a given correlation function
$C^{(\mathrm{1})}$ or $C^{(\mathrm{2})}$.
The averages and the quadratic averages over the disorder,
$C^{(\mathrm{1})}$ and $C^{(\mathrm{2})}$ are (see Appendix B for details):
\begin{eqnarray}
\label{THEOa0}
C^{(\mathrm{1})}&=&C^{(\mathrm{F})}, \quad \mathrm{in~F}, \\
\label{THEOa01}
C^{(\mathrm{1})}&=& 0, \quad k ~ \mathrm{odd}, \quad \mathrm{in~SG}, \\
\label{THEOa02}
C^{(\mathrm{1})}&=&C^{(\mathrm{SG})}, \quad k ~ \mathrm{even}, \quad \mathrm{in~SG},
\end{eqnarray} 
and
\begin{eqnarray}
\label{THEOa03}
C^{(\mathrm{2})}&=&\left(C^{(\mathrm{F})}\right)^2, \quad \mathrm{in~F}, \\
\label{THEOa04}
C^{(\mathrm{2})}&=&\left(C^{(\mathrm{SG})}\right)^2, \quad \mathrm{in~SG},
\end{eqnarray} 
where for sufficiently large $N$ 
\begin{eqnarray}
\label{THEOh}
{{C}}^{(\Sigma)}&=&
{{C}}_0(\beta J_0^{(\Sigma)};\{Np_jct^{(\Sigma)}m^{(\Sigma)}+\beta h\})
\nonumber \\
&& \mathop{O}\left(\frac{1}{N^{\delta}}\right),
\end{eqnarray} 
where in turn ${{C}}_0(\beta J_0,\{\beta h_i\})$ is the correlation function of the pure 
model (\ref{H0}), and finally $\delta\geq 1$ only for $k=1$, while in general $0\leq \delta<1$ for $k>1$
and $\delta=0$ if $\gamma<3$ (see Sec. \ref{CorrS}). 

From Eqs. (\ref{THEOa03}) and (\ref{THEOa04}) for $k=1$, we note that
the Edward-Anderson order parameter \cite{EA}
$C^{(\mathrm{2})}=\overline{\media{\sigma}^2}=q_{EA}$ is 
equal to $(C^{(\mathrm{SG})})^2=(m^{(\mathrm{SG})})^2$ only in the SG phase, whereas 
in the F phase we have $q_{EA}=(m^{(\mathrm{F})})^2$.
Therefore, since $m^{(\mathrm{SG})}\neq m^{(\mathrm{F})}$,
$m^{(\mathrm{SG})}$ is not equal to $\sqrt{q_{EA}}$;
in our approach $m^{(\mathrm{SG})}$ represents a sort of spin glass order parameter \cite{Parisi}.
In general, our method is able to establish exactly the phase boundary P-F and P-SG,
but not the frontiers F-SG when both the order parameters give a non zero solution.
Furthermore, while Eq. (\ref{orderpar}) for $\Sigma=$F can be derived, for $\Sigma=$SG
it remains only a plausible ansatz (see discussion at the end of the Sec. VII A of the Ref. \cite{SW}).
Note however that, at least for lattices $\mathcal{L}_0$
having only loops of even length, the stable P region is always that 
corresponding to a P-F phase diagram, so that in the P region
the correlation functions must be calculated only
through Eqs. (\ref{THEOa0}) and (\ref{THEOa03}).

As an immediate consequence of Eq. (\ref{THEOa}) we get the susceptibility 
$\tilde{\chi}^{(\Sigma)}$ of the model 
(throughout the paper we will use only the dimensionless 
definition of the susceptibility)

\begin{widetext}
\begin{eqnarray}
\label{THEOchie}
\tilde{\chi}^{(\Sigma)}\defi\frac{\partial m^{(\Sigma)}}{\partial(\beta h)}= 
\frac{\sum_ip_i\sum_j\tilde{\chi}_{0;i,j}\left(\beta J_0^{(\Sigma)};\{Np_lct^{(\Sigma)}m^{(\Sigma)}+\beta h\}\right)}
{1-ct^{(\Sigma)}N\sum_{i,j}\tilde{\chi}_{0;i,j}
\left(\beta J_0^{(\Sigma)};\{Np_lct^{(\Sigma)}m^{(\Sigma)}+\beta h\}\right)p_ip_j},
\end{eqnarray}
\end{widetext}


where $\tilde{\chi}_{0;i,j}$ stands for the two-points connected correlation function
of the pure model
\begin{eqnarray}
\label{THEOC2a}
\tilde{\chi}_{0;i,j}\defi 
\media{\sigma_i\sigma_j}_0-\media{\sigma_i}_0\media{\sigma_j}_0.
\end{eqnarray}
Note that, as it is evident in all the above equations, even when the pure model in the presence of a uniform external
field is translational invariant, for any non zero value of the order parameter $m^{(\Sigma)}$,
the disordered model is no longer translational invariant.
Note in particular that $\tilde{\chi}^{(\Sigma)}$ refers to the weighted
order parameter (\ref{orderpar}) so that it does not coincide
with the usual unweighted sum of the connected correlation functions.
In fact, from Eq. (\ref{THEOa}) it follows that 
\begin{eqnarray}
\label{THEOchie1}
\tilde{\chi}^{(\Sigma)}= 
\sum_{i,j}p_i\left[\overline{\media{\sigma_i\sigma_j}^{l_\Sigma}-
\media{\sigma_i}^{l_\Sigma}\media{\sigma_j}^{l_\Sigma}}\right].
\end{eqnarray}


\subsection{Critical surface (thermal and percolative)}
\label{Critical}



Note that, for $\beta$ sufficiently small (see later), 
Eq. (\ref{THEOa}) has always the solution $m^{(\Sigma)}=0$ and, furthermore,
if $m^{(\Sigma)}$ is a solution, $-m^{(\Sigma)}$ is a solution as well. 
From now on, if not explicitly said, 
we will refer only to the positive (possibly zero) solution,
the negative one being understood.
A solution $m^{(\Sigma)}$ of Eq. (\ref{THEOa}) is stable (but in general not unique) if
\begin{eqnarray}
\label{THEOO}
&&ct^{(\Sigma)}N\times  \\
&& \sum_{i,j}\tilde{\chi}_{0;i,j}
\left(\beta J_0^{(\Sigma)};\{Np_lct^{(\Sigma)}m^{(\Sigma)}+\beta h\}\right)p_ip_j<1. \nonumber
\end{eqnarray}
From Eq. (\ref{THEOa}) or from Eq. (\ref{THEOO}) we see that, in the thermodynamic limit, 
the critical surface crossing which the system passes from a P
region to a non P region satisfies  
\begin{eqnarray}
\label{THEOcrit}
ct_c^{(\Sigma)}N \sum_{i,j}\tilde{\chi}_{0;i,j}
\left(\beta_c^{(\Sigma)} J_0^{(\Sigma)};0\right)p_ip_j=1. 
\end{eqnarray}
Eq. (\ref{THEOcrit}) gives the critical surface of the model in the plane $(\beta,c)$ as a function
of $p$ and the other parameters of the model ($d\mu_0$, $d\mu$, $\{p_i\}$).

\subsubsection{Critical Temperature}
For a given value of $c$ Eq. (\ref{THEOcrit}) provides the critical temperature. 
From Eq. (\ref{THEOa}) it is immediate to recognize that for $J_0\neq 0$
\begin{eqnarray}
\label{THEOcrit2}
\beta_c^{(\Sigma)} <\beta_{c0}^{(\Sigma)},
\end{eqnarray}
while $\beta_c^{(\Sigma)} =\beta_{c0}^{(\Sigma)}$ for $J_0=0$,
where $\beta_{c0}^{(\Sigma)}$ is the critical temperature of the pure model with coupling $J_0^{(\Sigma)}$. 
It is clear that, when $(\mathcal{L}_0,\Gamma_0)$ is not translational invariant,
there exists an optimal choice of labeling the sites $i=1,\ldots,N$, which gives the lowest 
$\beta_{c}^{(\Sigma)}$, and that corresponds to the choice that maximizes the functional $F_\beta(\{p_i\})$, where  
\begin{eqnarray}
\label{THEOcrit4}
F_\beta(\{p_i\})\defi cN\sum_{i,j}\tilde{\chi}_{0;i,j}
\left(\beta J_0^{(\Sigma)};0\right)p_ip_j. \nonumber
\end{eqnarray} 
We will come back to this interesting issue in Sec. \ref{design}.

\subsubsection{Percolation threshold;
\label{Perco}
clustering versus percolation threshold}
The theory can be projected toward the limit $\beta\to\infty$ where for $\Sigma=$F 
we get an effective percolation theory.
Here the region P corresponds to the region in which,
in the thermodynamic limit, the parameters $(c,c_0,p)$
are such that no giant connected component exists ($m^{(F)}=0$).
Note in particular that, if $c_{0c}$ is the percolation threshold of 
the initial graph $(\mathcal{L}_0,\Gamma_0)$ (if $c_{0c}$ does not exist we can set formally $c_{0c}=\infty$)
in order to remain in the region P, 
the connectivity $c_0^{(p)}=c_0(1-p)$ of the graph obtained from the graph
$(\mathcal{L}_0,\Gamma_0)$ in which each link has been removed at random with probability $p$,
must satisfy $c_0(1-p)\leq c_{0c}$, otherwise a giant connected component already exists
(and the stability condition (\ref{THEOO}) at $\beta\to\infty$ with $m^{(F)}=0$ is violated).
From Eq. (\ref{THEOcrit}) in the thermodynamic limit it follows 
the equation for the percolation threshold $c_c$ as a function of $p$ 
\begin{eqnarray}
\label{THEOcrit5a}
&& c_cN \sum_{i,j}\tilde{\chi}_{0;i,j}\left(\tanh^{-1}(1-p);0\right)p_ip_j=1, \nonumber \\
&&\mathrm{with}\quad c_0(1-p)\leq c_{0c},
\end{eqnarray}
where we have used the fact that $\lim_{\beta\to\infty}\tanh(\beta J_0^{(\mathrm{F})})=\tanh(1-p)$.
Alternatively, Eq. (\ref{THEOcrit5a}) can be rewritten in terms of only graph elements as
\begin{eqnarray}
\label{THEOcrit5}
c_cN \sum_{i,j}\left(\delta_{i,j}+\mathcal{N}_{0;i,j}^{(p)}\right)p_ip_j=1, \quad c_0(1-p)\leq c_{0c},
\end{eqnarray}
where $\mathcal{N}_{0;i,j}^{(p)}=1$ if, in the graph $(\mathcal{L}_0,\Gamma_0)$
from which each link has been removed at random with probability $p$,
between the vertex $i$ and the vertex $j$ there exists at least a path of links, 
and $\mathcal{N}_{0;i,j}^{(p)}=0$ otherwise.

Given $p$, if the condition $\quad c_0(1-p)\leq c_{0c}$ is not satisfied, then
a giant connected component is present and we can set $c_c=0$.
It is interesting to see in more details the case in which we choose $c=c_0p$ so
that, as we vary $p$, the total connectivity is fixed and equal to $c_0$ (the ``rewired'' small-world).
This study is important since it leads us to understand how the presence of short loops affects
diffusion processes on general networks. In particular, a strong interest
regards the question: \textit{``In the presence of short loops how does 
the percolation threshold change''?}.
If we set $c=c_0p$, from Eq. (\ref{THEOcrit5}) we get the percolation threshold $c_{0c}$ as
a function of the rewiring parameter $p$
\begin{eqnarray}
\label{THEOcrit5b}
c_{0c}^{(p)}pN \sum_{i,j}\left(\delta_{i,j}+\mathcal{N}_{0;i,j}^{(p)}\right)p_ip_j=1, \quad c_0^{(p)}(1-p)\leq c_{0c}.
\end{eqnarray}
From Eq. (\ref{THEOcrit5b}) we see that $p$ has two effects on $c_{0c}^{(p)}$: 
the pref-actor $p$ in the lhs of Eq. (\ref{THEOcrit5b}) tends to decrease $c_{0c}^{(p)}$, 
while the other tends to decrease $\mathcal{N}_{0;i,j}^{(p)}$ and then to increase $c_{0c}^{(p)}$. 
However, as we shall see soon, in general $c_{0c}^{(p)}$ decreases with $p$ due to the
general mechanism according to which clustering diminishes the percolation threshold. 

\begin{figure}[tbh]
\epsfxsize=75mm \centerline{\epsffile{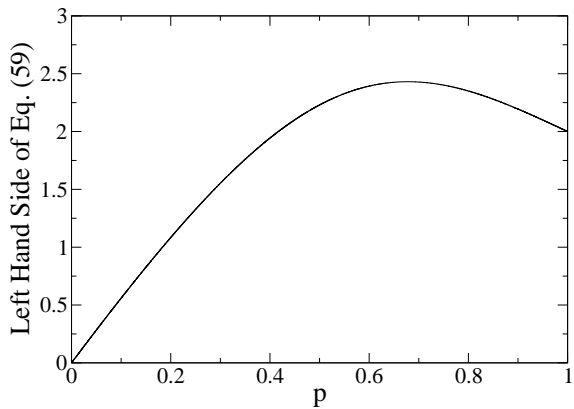}}
\caption{Plot of the Left Hand Side of Eq. (\ref{THEOcrit5fc1}) ($(\mathcal{L}_0,\Gamma_0)$ is a set of
disjoint triangles) as a function of the dilution probability $p$.}
\label{fig0}
\end{figure}

A special case is the one in which $p_i\equiv 1/N$;
\textit{i.e.}, the classical small-world (no heterogeneity).
In this case Eq. (\ref{THEOcrit5a}) simplifies as
\begin{eqnarray}
\label{THEOcrit5c}
c_{0c}^{(p)}p\tilde{\chi}_0(\tanh^{-1}(1-p);0)=1, \quad c_0^{(p)}(1-p)\leq c_{0c}.
\end{eqnarray}
So, for example, if $(\mathcal{L}_0,\Gamma_0)$ is the 
Erd$\mathrm{\ddot{o}}$s-R$\mathrm{\acute{e}}$nyi random graph \cite{Classical} (in the canonical representation),
with mean connectivity $c_{0}$, from (valid in the P region) 
\begin{eqnarray}
\label{THEOcrit5e}
\tilde{\chi}_0(\beta J_0;0)=\frac{1}{1-c_0\tanh(\beta J_0)},
\end{eqnarray}
to be inserted in Eq. (\ref{THEOcrit5c}), 
we get back obviously the well known percolation threshold $c_{0c}^{(p)}=1$, independently of the value of $p$.  
Depending on the problem, given $c_0<c_{0c}$, in general one can be more interested in reading Eq. (\ref{THEOcrit5a})
either as an equation for $p$ or for $c$. 
We can consider for example the case in which $(\mathcal{L}_0,\Gamma_0)$
is an ensemble of arbitrary disconnected finite clusters (dimers, triangles, \dots, or mixtures of them),
for which there is no percolation threshold (or formally $c_{c0}=\infty$).
For example, for a set of $N/2$ disconnected dimers ($c_0=1$), $N/3$ disconnected triples 
($c_0=1\times 2/3+2\times 1/3$), $N/3$ 
disconnected triangles ($c_0=2$), $N/4$
disconnected squares ($c_0=2$), and $N/5$
disconnected pentagons ($c_0=2$),
we have respectively:
\begin{eqnarray}
\label{THEOcrit5f}
\tilde{\chi}_0(\beta J_0;0)=\frac{2e^{\beta J_0}}{e^{\beta J_0}+2e^{-\beta J_0}} \quad (\mathrm{dimers}),
\end{eqnarray}
\begin{eqnarray}
\label{THEOcrit5fb}
\tilde{\chi}_0(\beta J_0;0)=\frac{1}{3}\frac{9e^{2\beta J_0}+2+2e^{-2\beta J_0}}{e^{2\beta J_0}+2+e^{-2\beta J_0}}
\quad (\mathrm{triples}),
\end{eqnarray}
\begin{eqnarray}
\label{THEOcrit5fc}
\tilde{\chi}_0(\beta J_0;0)=\frac{3e^{3\beta J_0}+e^{-\beta J_0}}{e^{3\beta J_0}+3e^{-\beta J_0}}
\quad (\mathrm{triangles}),
\end{eqnarray}
\begin{eqnarray}
\label{THEOcrit5fd}
\tilde{\chi}_0(\beta J_0;0)=\frac{4e^{4\beta J_0}+4}{e^{4\beta J_0}+7}
\quad (\mathrm{squares}),
\end{eqnarray}
\begin{eqnarray}
\label{THEOcrit5fe}
\tilde{\chi}_0(\beta J_0;0)=\frac{5e^{5\beta J_0}+11e^{\beta J_0}}{e^{5\beta J_0}+15e^{\beta J_0}}
\quad (\mathrm{pentagons}).
\end{eqnarray}
If we consider the case with no heterogeneity $p_i\equiv 1/N$, from
Eqs. (\ref{THEOcrit5f})-(\ref{THEOcrit5fe}) plugged in Eq. (\ref{THEOcrit5a})
for $p=0$, 
we get respectively the following percolation thresholds $c_c$:
\begin{eqnarray}
\label{THEOcritc5f}
c_{c}=1/2 \quad (\mathrm{dimers}),
\end{eqnarray}
\begin{eqnarray}
\label{THEOcritc5fb}
c_{c}=1/3 \quad (\mathrm{triples}),
\end{eqnarray}
\begin{eqnarray}
\label{THEOcritc5fc}
c_{c}=1/3 \quad (\mathrm{triangles}),
\end{eqnarray}
\begin{eqnarray}
\label{THEOcritc5fd}
c_{c}=1/4 \quad (\mathrm{squares}),
\end{eqnarray}
\begin{eqnarray}
\label{THEOcritc5fe}
c_{c}=1/5 \quad (\mathrm{pentagons}).
\end{eqnarray}
and, in general, for polygons of $m\geq 3$ sides ($c_0=2$)
\begin{eqnarray}
\label{THEOcritc5fe}
c_{c}=1/m \quad (\mathrm{polygons~of~}m\mathrm{~sides}).
\end{eqnarray}
Notice that the clustering coefficient 
for dimers and triples is zero, and
for closed polygons of $m$ sides decreases with $m$. 
Of course one recovers that $c_{c}=1/m \to 0$ for $m\to\infty$
since an ensemble of $N/m$ disconnected polygons of length $m$
for $m=N\to\infty$ becomes equivalent to a closed chain for which
we already know that $c_{c}=0$.
Eqs. (\ref{THEOcrit5f}-\ref{THEOcrit5fe}) can be used in general also for $p>0$. 
So, for example,
from Eq. (\ref{THEOcrit5c}) and (\ref{THEOcrit5fc}), by using the replacement $\beta J_0\to\tanh^{-1}(1-p)$,
we get the equation for the percolation threshold $p_c$ of an ensemble of disconnected triangles ($c_0=2$)
from which each link has been removed with probability $p$ ($c_0^{(p)}=2p$) and ``rewired'' as a ``long-range''
link: 
\begin{eqnarray}
\label{THEOcrit5fc1}
2p\frac{3e^{3\tanh^{-1}(1-p)}+e^{-\tanh^{-1}(1-p)}}{e^{3\tanh^{-1}(1-p)}+3e^{-\tanh^{-1}(1-p)}}=1.
\end{eqnarray}
In Fig. (\ref{fig0}) we plot the lhs of Eq. (\ref{THEOcrit5fc1}) as a function of $p$. 
Eq. (\ref{THEOcrit5fc1}) is solved for $p_c=0.183406$.\vspace{1.00cm}

Let us come back now to the general heterogeneous case.
From Eq. (\ref{THEOcrit5b}) we see that, 
given two regular graphs $(\mathcal{L}_0,\Gamma_0)$ and $(\mathcal{L}_0,\Gamma_0')$, 
both having the same average connectivity $c_0=c_0'$ (so that $|\Gamma_0'|=|\Gamma_0|$), 
between $(\mathcal{L}_0,\Gamma_0)$ and $(\mathcal{L}_0,\Gamma_0')$,
the sum in the lhs of Eq. (\ref{THEOcrit5b}) will be greater for the
graph having the smaller clustering coefficient, 
which in turn will result in a lower value for $c_{0c}^{(p)}$.
In fact, 
given a vertex $i$ and its local connectivity $c_0(i)$, the smaller is the clustering coefficient
around the vertex $i$,
the larger will be the number of different vertices $j$ connected to $i$
(while when the clustering coefficient is large a same vertex $j$ will be reached from the vertex $i$ 
by many different paths),  
so that $\sum_j\mathcal{N}_{0;i,j}$ will be greater
which in turn will give rise, via Eq. (\ref{THEOcrit5b}), to a smaller clustering coefficient.
And similarly for  
$\mathcal{N}_{0;i,j}^{(p)}$ for any given $p$.  
In conclusion, as already discussed in \cite{Newman,Miller,Gleeson1},
clustering increases the percolation threshold.
Equation (\ref{THEOcritc5fe}) for the polygons represents
a clear example of this mechanism for the particular choice $c_0=2$, $p=0$
and $p_i\equiv 1/N$. 

We conclude this Section with a remark on the recent 
methods used by Newman \cite{HyperNewman} and Gleeson \cite{HyperGleeson}
by which families of clustered networks are introduced and
analytically exactly solved by generating function techniques.
Although these networks have a finite clustering coefficient,
they can still be mapped to effective tree-like graphs.
So, for example, for the ensemble of disconnected finite clusters
as the ones we have analyzed in Eqs. (\ref{THEOcrit5f})-(\ref{THEOcritc5fe}),
we could also use the method \cite{HyperNewman} to solve the percolation
problem \footnote{
\textit{e.g.}, for the ensemble of disconnected
triangles, by using the same formalism of \cite{HyperNewman},
it is easy to see that by choosing $p_{s,t}=p_s\delta_{t,1}$,
$p_s$ being Poissonian with mean $\mu=c$, we reach Eq. (\ref{THEOcritc5fc}).}
, but not for example the case in which $(\mathcal{L}_0,\Gamma_0)$
is a $d_0$ dimensional lattice. In fact, the main condition which allows
the methods \cite{HyperNewman} or \cite{HyperGleeson} to be applied is the
absence of overlaps among the module-elements (links, or triangles, or any kind
of finite cluster) while, for example, in a two-dimensional lattice
we have always overlap among the square ``module-elements''. 

\subsection{Critical behavior}
In this section we prove that the critical behavior of an arbitrary heterogeneous
graph as defined through Eq. (\ref{hidden}), which in particular includes scale-free graphs, 
is robust with respect to the addition of any graph $(\mathcal{L}_0,\Gamma_0)$,
provided $(\mathcal{L}_0,\Gamma_0)$ is not in turn a heterogeneous graph.
In \cite{SW} we have shown this result for the homogeneous small-world model
corresponding to the case $p_i\equiv 1/N$. More precisely the critical behavior for $p_i\equiv 1/N$ and $p=0$
has been shown to be classical mean field for $t_0^{(\mathrm{F})}\geq 0$, while
for $t_0^{(\mathrm{F})}< 0$ first-order phase transitions are also possible (see also \cite{SWMC}).
Here we will restrict the analysis only to the case $t_0^{(\mathrm{F})}\geq 0$.  
First of all from Eqs. (\ref{THEOchie}) and (\ref{THEOcrit}) we observe immediately that  
the critical exponents for the susceptibility, above and below the critical temperature, 
are both equal to 1. Note in particular that above the critical temperature 
and zero external field the susceptibility can be written in the simpler form
\begin{eqnarray}
\label{THEOchie2}
\tilde{\chi}^{(\Sigma)}= 
\frac{\sum_ip_i\sum_j\tilde{\chi}_{0;i,j}\left(\beta J_0^{(\Sigma)};\{0\}\right)}
{1-t^{(\Sigma)}/t_c^{(\Sigma)}}.
\end{eqnarray}
Let us now turn to the analysis of the order parameter near the critical point.
For $J_0=0$, \textit{i.e.}, for the pure static model, we have 
\begin{eqnarray}
\label{THEOp}
m_{0i}(\beta J_0^{(\Sigma)};\{\beta h_j\})=
\tanh(\beta h_i)
\end{eqnarray} 
so that the self-consistent equation (\ref{THEOa}) strongly simplifies in
\begin{eqnarray}
\label{THEOp1}
m^{(\Sigma)}=g(m^{(\Sigma)})
\end{eqnarray}
where 
\begin{eqnarray}
\label{THEOp4b}
g(m^{(\Sigma)})\defi \sum_i \tanh(Np_ict^{(\Sigma)}m^{(\Sigma)})p_i .
\end{eqnarray}
The critical behavior of pure static model, 
\textit{i.e.} with $J_0=0$, for the scale-free choice (\ref{stat})-(\ref{stat1}),
has been studied in the Ref. \cite{KimStat}. 
Let us focus on the P-F transition.
For $\Sigma=$F, Eq. (\ref{THEOp1}) is equal to Eq. (21) of \cite{KimStat}.
We recall that, due to the power-law character of the distribution $\{p_j\}$, we cannot derive the correct
critical behavior by simply expanding in the sum in $g(m^{(\mathrm{F})})$ term by term for small $m^{(\mathrm{F})}$. 
As shown in \cite{KimStat}, it is necessary to keep track of all the terms of the sum present
in $g(m^{(\mathrm{F})})$. This is done by evaluating the sum with the corresponding integral
which gives rise to a singular term proportional to $(m^{(\mathrm{F})})^{\gamma-2}$ plus regular terms
proportional to $(m^{(\mathrm{F})})$, $(m^{(\mathrm{F})})^{3}$ and so on.
As a consequence, when we solve the self-consistent equation to leading order in $m^{(\mathrm{F})}$, 
if $T$ and $\tau$ indicate the temperature and the reduced temperature, respectively,
we get the well known anomalous mean-field behavior: 
$m^{(\mathrm{F})}\sim\mathop{O}(\tau^{1/2})$ (\textit{i.e.} classical mean-field) for $\gamma>5$,  
$m^{(\mathrm{F})}\sim\mathop{O}(\tau^{1/(\gamma-3)})$ for $3<\gamma<5$, and  
$m^{(\mathrm{F})}\sim\mathop{O}(T^{-(\gamma-2)/(3-\gamma)})$ for $2<\gamma<3$.  
Note that the critical behavior of the order parameter $m^{(\mathrm{F})}=\sum_i\overline{\mediaT{\sigma_i}}p_i$ 
is different from the unweighted 
one defined as $\overline{m}=\sum_i\overline{\mediaT{\sigma_i}}/N$ when $2<\gamma<3$. 
In such a case from Eq. (\ref{THEOh}) one can use 
$\overline{m}\sim t^{(\mathrm{F})}m^{(\mathrm{F})}$ from which it follows that 
$\overline{m}\sim\mathop{O}(T^{-1/(3-\gamma)})$ for $2<\gamma<3$. 

Let now be $J_0\neq 0$. 
It is clear that if the graph $(\mathcal{L}_0,\Gamma_0)$ is in turn
a pure scale free graph with exponent $\gamma'$, then the joined network
will have an anomalous critical behavior characterized by the minimum between
$\gamma$ and $\gamma'$. Less obvious is to understand what happens if $(\mathcal{L}_0,\Gamma_0)$
has a finite dimensional structure or some special topology with short loops.
In particular we can pose the question: does the critical behavior change
by adding, via short loops, many paths between far spins, or may the critical exponent for $m$ depend on $J_0$? 
Let us consider the self-consistent equation (\ref{THEOa}) in general.     
The exact expression of $m_{0i}(\beta J_0^{(\mathrm{F})};\{\beta h_j\})$ for a generic
non homogeneous external field $\{h_j\}$ represents a formidable task. 
Note that, as above mentioned, to analyze the critical behavior we cannot expand for small fields $\{\beta h_j\}$.
We can however perform an expansion to the lowest order in $t_0^{(\mathrm{F})}=\tanh(\beta J_0^{(\mathrm{F})})$.
It is then easy to see that, for $\{h_j\}\neq 0$, at the order $\mathop{O}(t_0^{(\mathrm{F})})$ we have 
\begin{eqnarray}
\label{THEOp2}
&& m_{0i}(\beta J_0;\{\beta h_j\})=\tanh(\beta h_i) +\nonumber \\
&& t_0^{(\mathrm{F})}\left[1-\tanh^2(\beta h_i)\right]
\sum_{j\in\mathcal{N}_0(i)}\tanh(\beta h_j),
\end{eqnarray} 
where $\mathcal{N}_0(i)$ is the set of the first neighbors of the vertex $i$ in the graph $(\mathcal{L}_0,\Gamma_0)$.
It must be said that without the condition $\{h_j\}\neq 0$ Eq. (\ref{THEOp2}) might be wrong since
the lowest non zero terms in $t_0^{(\mathrm{F})}$ would involve closed paths of at least length 3, while 
Eq. (\ref{THEOp2}) contains only paths of length 1.
More precisely, near the critical point, due to the fact that the fields $\{h_j\}$ are infinitesimal but not zero,
we can neglect higher order corrections in $t_0^{(\mathrm{F})}$.
By plugging Eq. (\ref{THEOp2}) into Eq. (\ref{THEOa}) for $\Sigma=$F we have
\begin{eqnarray}
\label{THEOp3}
&& m^{(\mathrm{F})}=g(m^{(\mathrm{F})})+\Delta_1(m^{(\mathrm{F})})+\Delta_2(m^{(\mathrm{F})}),
\end{eqnarray}
where we have introduced
\begin{eqnarray}
\label{THEOp4}
\Delta_1(m^{(\mathrm{F})})\defi 
t_0^{(\mathrm{F})}\sum_i p_i\sum_{j\in\mathcal{N}_0(i)}\tanh(Np_jct^{(\mathrm{F})}m^{(\mathrm{F})}),
\end{eqnarray}
\begin{eqnarray}
\label{THEOp4c}
\Delta_2(m^{(\mathrm{F})})&\defi& 
t_0^{(\mathrm{F})}\sum_i p_i \tanh^2(Np_ict^{(\mathrm{F})}m^{(\mathrm{F})}) \nonumber \\
&& \times \sum_{j\in\mathcal{N}_0(i)}\tanh(Np_jct^{(\mathrm{F})}m^{(\mathrm{F})}).
\end{eqnarray}
both to be compared with the $J_0$ independent term $g(m^{(\mathrm{F})})$.
Let us analyze the bigger contribution $\Delta_1(m^{(\mathrm{F})})$ and
let us focus on the simpler cases in which the graph $(\mathcal{L}_0,\Gamma_0)$
has a fixed connectivity $|\mathcal{N}_0(i)|\equiv c_0$, where $|\mathcal{N}_0(i)|$ stands for
the cardinality of the set $\mathcal{N}_0(i)$.  
Let us suppose first that $c_0=1$ (\textit{i.e.}, $(\mathcal{L}_0,\Gamma_0)$ is an ensemble of dimers).
In general, given any normalized distribution $p_i\geq 0$, different from the homogeneous one, 
and any function $f(x)\geq 0$
increasing with $x$, the following property holds
\begin{eqnarray}
\label{THEOp5}
\sum_i p_i f(p_{j_0(i)})< \sum_i p_i f(p_i), 
\end{eqnarray}
where $j_0(i)$ stands for the first single neighbor of $i$ in $(\mathcal{L}_0,\Gamma_0)$,
while 
\begin{eqnarray}
\label{THEOp5b}
\sum_i p_i f(p_{j_0(i)})= \sum_i p_i f(p_i), 
\end{eqnarray}
only for $p_i\equiv 1/N$.
Note that by definition $i\rightarrow j_0(i)$ is a bijection on $\mathcal{L}_0$ and that
$j_0(i)\neq i$. We can however formally enlarge the definition of $j_0(i)\neq i$ 
to include also the case $j_0(i)=i$ (a self-link).
The inequality (\ref{THEOp5}) tells us that when we choose $j_0(i)=i$ we get an optimal
overlap between the distribution $\{p_i\}$ and the function $f(\cdot)$. 
For the general case $|\mathcal{N}_0(i)|\equiv c_0\geq 1$, given a vertex $i$, we can
enumerate the $c_0$ neighbors of $i$ as $j_0^{(1)}(i),\ldots,j_0^{(c_0)}(i)$. Each upper index $l=1,\ldots,c_0$ 
represents an oriented axes so that, for each $l=1,\ldots,c_0$, the function 
$j_0^{(l)}(i)$ is a bijection on $\mathcal{L}_0$.
For example, if $(\mathcal{L}_0,\Gamma_0)$ is the one dimensional ring,
we have the two bijections $j_0^{(1)}(i)=i+1$ and $j_0^{(2)}(i)=i-1$.  
By applying Eq. (\ref{THEOp5}) to each oriented axes we then get 
\begin{eqnarray}
\label{THEOp6}
&&\sum_i p_i \sum_{j\in\mathcal{N}_0(i)} f(p_j)=
\sum_{l=1}^{c_0}\sum_i p_i f(p_{j_0^{(l)}(i)})\nonumber \\
&& < c_0\sum_i p_i f(p_i).  
\end{eqnarray}
By using Eq. (\ref{THEOp6}) to our case with $f(x)=\tanh(x)$ and for $t_0^{(\mathrm{F})}>0$
we see that for $m^{(\mathrm{F})}>0$ we have always
\begin{eqnarray}
\label{THEOp7}
0< \Delta_1(m^{(\mathrm{F})})<t_0^{(\mathrm{F})}c_0 g(m^{(\mathrm{F})}).
\end{eqnarray}
In turn $0< \Delta_2(m^{(\mathrm{F})})<\Delta_1(m^{(\mathrm{F})})$ and,
furthermore, as already mentioned 
- near the critical point - higher order corrections in $t_0^{(\mathrm{F})}$ will be all lower
than the first term proportional to $t_0^{(\mathrm{F})}$.
In conclusion, for $t_0^{(\mathrm{F})}>0$, from Eq. (\ref{THEOa}) and the above inequalities, we get 
\begin{eqnarray}
\label{THEOp8}
g(m^{(\mathrm{F})})< m^{(\mathrm{F})}&<&\left(1+t_0^{(\mathrm{F})}c_0 \right)g(m^{(\mathrm{F})})+\nonumber \\
&&\mathop{O}(t_0^2)g(m^{})+\mathop{o}(g(m^{(\mathrm{F})}),
\end{eqnarray}
where $\mathop{o}(g(m^{(\mathrm{F})})$ stands for corrections smaller than $g(m^{(\mathrm{F})})$.
In general, for $t_0$ finite, it is possible to prove that 
\begin{eqnarray}
\label{THEOp8g}
g(m^{})< m^{}<\tilde{\chi}_0\left(\beta J_0^{(\mathrm{F})};0\right)g(m^{})+\mathop{o}(g(m^{})),
\end{eqnarray} 
where $\tilde{\chi}_0(\beta J_0^{(\mathrm{F})};0)$ is the susceptibility of
the pure model (\ref{H0}) with coupling $J_0^{(\mathrm{F})}$ and $h_i\equiv 0$.
The proof is given in Appendix D. 
Since near the critical point, in the region $\beta_{c0}^{(F)}>\beta>\beta_c$, it is
$\tilde{\chi}_0(\beta J_0^{(\mathrm{F})};0)<\infty$, we see that
Eq. (\ref{THEOp8g}) implies that the critical behavior of Eq. (\ref{THEOa})
remains always that 
corresponding to the term $g(m^{})$, \textit{i.e.} as if it were $J_0=0$.
We can finally consider the case in which $(\mathcal{L}_0,\Gamma_0)$ is a
Poissonian graph (Erd$\mathrm{\ddot{o}}$s-R$\mathrm{\acute{e}}$nyi random graph
in the canonical representation) with mean connectivity $c_0$. To this aim
we can start from the fully connected graph and remove from it randomly
each of its $N(N-1)/2$ links with a probability $p=1-c_0/N$; the resulting graph
will be our Poissonian graph with mean connectivity $c_0$. 
Since we have already proved
that when $(\mathcal{L}_0,\Gamma_0)$ is the fully connected graph 
with a couplings $\mathop{O}(1/N)$ the critical behavior 
remains equal to that of the model with $J_0=0$ (the effective
couplings in this case being given by $\tan(\beta J_0^{(\mathrm{F})})=\tanh(\beta J_0)c_0/N$), 
we conclude that also for a Poissonian graph the critical behavior
of the small-world model remains the same as it were $J_0=0$.

For the P-SG transition of the model with $J_0=0$ we can evaluate
the weighted Edward-Anderson order parameter as $q_{EA}=(m^{(\mathrm{SG})})^2$.
Since the critical behavior of $m^{(\mathrm{SG})}$ is identical to that
of $m^{(\mathrm{F})}$ we get that the critical exponent for $q_{EA}$ is
is simply given by twice the $\gamma$ dependent critical exponent for $m^{(\mathrm{F})}$
that we have discussed before. This result is in contrast with the one of the
Ref. \cite{KimStat} for the region $4<\gamma<5$. The source of such a contrast
might be related to the already mentioned fact that
Eq. (\ref{orderpar}) for $\Sigma=$SG remains only a plausible ansatz.
We do not discuss here further this issue, but we stress that, whatever be
the critical P-SG behavior of the model with $J_0=0$, by applying
the same procedure as above done for the case P-F, we arrive at the conclusion that
also the critical P-SG behavior is infinitely robust with respect to the addition
of any non heterogeneous graph $(\mathcal{L}_0,\Gamma_0)$. 
Same conclusions hold of course also for the critical exponent of
the order parameter of the percolation problem.

\subsection{Correlation functions}
\label{CorrS}
Another remarkable consequence of our theory comes from Eq. (\ref{THEOh}).
We see in fact that, in the thermodynamic limit, any correlation function
of the model, at least for $\gamma>3$, fits with the correlation function of the pure model
but immersed in an effective field that is exactly 
zero in the P region and zero external field ($\{h=0\}$).
In other words, in terms of correlation functions, 
in the P region, the small-world model and the pure model are indistinguishable
(modulo the transformation $J_0\to J_0^{(\mathrm{SG})}$ for $\Sigma=$SG).
Note however that this assertion holds only for a given correlation function 
calculated in the thermodynamic limit. In fact, the corrective $\mathop{O}(1/N^{\delta})$ term
appearing in the rhs of Eq. (\ref{THEOh}) cannot be neglected when
we sum the correlation functions over all the sites $i\in\mathcal{L}_0$,
as to calculate the susceptibility; yet it is just
this corrective $\mathop{O}(1/N^{\delta})$ 
term that gives rise to the singularities of the model.
More precisely, for the two point connected correlation function defined as
\begin{eqnarray}
\label{THEOC2}
\tilde{\chi}_{i,j}^{(\Sigma)}\defi 
\overline{\media{\sigma_i\sigma_j}^{l_\Sigma}-\media{\sigma_i}^{l_\Sigma}\media{\sigma_j}^{l_\Sigma}},
\end{eqnarray}
where $^{l_\Sigma}=1,2$ for $\Sigma=$ F, SG, respectively, we have
\begin{widetext}
\begin{eqnarray}
\label{THEOC2b}
\tilde{\chi}_{i,j}^{(\Sigma)}&=&
\tilde{\chi}_{0;i,j}(\beta J_0^{(\Sigma)};\{Np_jct^{(\Sigma)}m^{(\Sigma)}+\beta h\})+
\nonumber \\
&& ct^{(\Sigma)}N 
\frac{\sum_l\tilde{\chi}_{0;i,l} (\beta J_0^{(\Sigma)};\{Np_qct^{(\Sigma)}m^{(\Sigma)}+\beta h\})p_l
\sum_np_n\tilde{\chi}_{0;n,j} (\beta J_0^{(\Sigma)};\{Np_qct^{(\Sigma)}m^{(\Sigma)}+\beta h\})}
{1-ct^{(\Sigma)}N\sum_{l,n}\tilde{\chi}_{0;l,n}
\left(\beta J_0^{(\Sigma)};\{Np_qct^{(\Sigma)}m^{(\Sigma)}+\beta h\}\right)p_lp_n},
\end{eqnarray}
where the dependence on $N$ in $\tilde{\chi}_{i,j}^{(\Sigma)}$ and $\tilde{\chi}_{0;i,j}$ are understood. 
In the homogeneous case $p_i\equiv 1/N$ Eq. (\ref{THEOC2b}) becomes
\begin{eqnarray}
\label{THEOC2c}
\tilde{\chi}_{i,j}^{(\Sigma)}&=&
\tilde{\chi}_{0;i,j}(\beta J_0^{(\Sigma)};Nct^{(\Sigma)}m^{(\Sigma)}+\beta h)+
\nonumber \\
&& \frac{ct^{(\Sigma)}}{N}
\frac{\sum_l\tilde{\chi}_{0;i,l} (\beta J_0^{(\Sigma)};ct^{(\Sigma)}m^{(\Sigma)}+\beta h)
\sum_n\tilde{\chi}_{0;n,j} (\beta J_0^{(\Sigma)};ct^{(\Sigma)}m^{(\Sigma)}+\beta h)}
{1-ct^{(\Sigma)}\tilde{\chi}_{0}\left(\beta J_0^{(\Sigma)};ct^{(\Sigma)}m^{(\Sigma)}+\beta h\right)},
\end{eqnarray}
which, when $(\mathcal{L}_0,\Gamma_0)$ is in turn homogeneous, reduces to \cite{SW}
\begin{eqnarray}
\label{THEOC2d}
\tilde{\chi}_{i,j}^{(\Sigma)}&=&
\tilde{\chi}_{0;i,j}(\beta J_0^{(\Sigma)};Nct^{(\Sigma)}m^{(\Sigma)}+\beta h)+
\frac{ct^{(\Sigma)}}{N}
\frac{\left[\tilde{\chi}_{0}\left(\beta J_0^{(\Sigma)};ct^{(\Sigma)}m^{(\Sigma)}+\beta h\right)\right]^2}
{1-ct^{(\Sigma)}\tilde{\chi}_{0}
\left(\beta J_0^{(\Sigma)};ct^{(\Sigma)}m^{(\Sigma)}+\beta h\right)}.
\end{eqnarray}
\end{widetext}

Eq. (\ref{THEOC2b}) is easily obtained by derivation of the mean-field equation 
(\ref{THEOa}) generalized to the case of an arbitrary external field $\{h_i\}$ (see Appendix E).
Eq. (\ref{THEOC2b}) clarifies the structure of the correlation
functions in general small-world models. In the rhs we have two terms: the former
is a distance-dependent (the distance, if any, defined in the graph $(\mathcal{L}_0,\Gamma_0)$) 
short-range term whose finite correlation length, for $T\neq T_{c0}^{(\Sigma)}$, 
makes it summable (over all the nodes $(i,j)$), the latter is instead a term which takes into account
the heterogeneity of the system - possibly power law like - also in the P phase, 
which turns out
to be summable thanks to a global $1/N^{\delta}$ factor, where $\delta$ is the exponent
appearing in Eq. (\ref{THEOh}). For the two point connected correlation
function, at least for the case in which $(\mathcal{L}_0,\Gamma_0)$ is a regular lattice, 
as will be clear in the next Section, $\delta$ takes the value:
\begin{eqnarray}
\label{delta}
\delta=\left\{
\begin{array}{l}
\frac{\gamma-3}{\gamma-1}, \quad \mathrm{for~} \gamma>3\\
0, \quad \mathrm{for~} 2<\gamma\leq 3. 
\end{array}
\right.
\end{eqnarray}
Once we perform the weighted
sums with the distribution $\{p_i\}$, both the terms in the rhs of Eq. (\ref{THEOC2b}) give a finite
contribution to the susceptibility.
It is in fact immediate to verify that by inserting Eq. (\ref{THEOC2b}) in Eq. (\ref{THEOchie1})
we get back Eq. (\ref{THEOchie}). 

We see here a novel fact: in scale free models 
in finite but large systems, correlations between two given spins can be power law like even above
the critical surface. Furthermore, we see from Eq. (\ref{delta})
that such phenomena become persistent even in the thermodynamic limit when
$\gamma\leq 3$. 
At this point, it is worth to compare these scenario,
with the scenarios one has in other systems.
By focusing only on the second term of the rhs of Eq. (\ref{THEOC2b})
we find the following.
In finite $d$-dimensional models, according to the Ornstein-Zernike form \cite{Landau},
at any $T$ but the critical one $T_c$, one has exponentially small correlations, while at $T_c$
the correlation function decays as a power law with the distance with an exponent $d-2+\eta$,
where $\eta$ is the critical exponent of the correlation length.
Roughly speaking, this implies that in finite dimensional models there are
essentially two possible correlations for near (\textit{i.e.} at a distance $\mathop{O}(1)$ at $T\neq T_c$) 
and far (\textit{i.e.} at a distance $\mathop{O}(N)$ at $T=T_c$) spins, 
with values $\mathop{O}(1)$ and $\mathop{O}(1/N^{d-2+\eta})$, respectively, and 
the total number of such couples of spins are $\mathop{O}(N)$ and $\mathop{O}(N^2)$, respectively.
In the fully connected model with a coupling $\mathop{O}(1/N)$, or in classical
random graphs, or in homogeneous small-world models, at any $T$ the correlation function decays
instead as $1/N$ for any couple of the $N(N-1)/2$ spins, with no spatial dependence
(the correlation length goes to infinity).
As will be clear in the next Section,
in heterogeneous small-world networks with a power law degree distribution $k^{-\gamma}$, at any $T$ 
we can instead distinguish three families of correlations:
given two spins that are both far from an hub, they have correlations 
$\mathop{O}(1/N)$, and the total number of such couples of spins is of the order $\mathop{O}(aN^2)$
with $a<1$; 
given two spins, one of which is an hub and the other not, they have correlations
$\mathop{O}(1/N^{(\gamma-2)/(\gamma-1)})$, and the number of such couples of spins is
$\mathop{O}(b N)$, where $b$ is a decreasing function of $1/\gamma$; 
finally, given two spins which are both an hub, they have correlations
$\mathop{O}(1/N^{(\gamma-3)/(\gamma-1)})$, and the number of such couples of spins is $\mathop{O}(1)$.

\section{Examples}
In Sec. \ref{Perco} we have seen some simple applications to the homogeneous case $(p_i\equiv 1/N)$.
Here we discuss some examples where we can apply, analytically, the general 
results of the previous Section to the heterogeneous case.
Since we have already solved the issued for the critical behavior
we will focus only on the critical surface and on the correlation functions. 

\subsection{Viana Bray on the scale-free graph}
In this case $J_0=0$ so that there is no additional graph, for historical reasons
we refer this as the Viana Bray model \cite{VianaBray} on the scale-free graph. 
This model was solved in \cite{KimStat}, and for the network version called ``configuration model''
(which is a network realization slightly different from the 
hidden variables network) the Ising model was already extensively studied
in \cite{GZ} almost one decade ago. 
Since $J_0=0$ for $m_0$ we can use
Eq. (\ref{THEOp}) from which in particular it follows that for $\beta<\beta_{c0}$
we have $\tilde{\chi}_{0;i,j}=\delta_{i,j}$.
By inserting this in Eq. (\ref{THEOcrit}) we get the critical surface $t_c^{(\Sigma)}$
\begin{eqnarray}
\label{THEOcritpure}
ct_c^{(\Sigma)}N \sum_i p_i^2=1,
\end{eqnarray}
which, for large $N$ under the choice (\ref{stat})-(\ref{stat1}) gives
\begin{eqnarray}
\label{THEOcritpure1}
ct_c^{(\Sigma)}\frac{(1-\mu)^2}{(1-2\mu)}(1-N^{2\mu-1})=1.
\end{eqnarray}
The critical surface given by Eq. (\ref{THEOcritpure})
coincides with the one found in \cite{KimStat}.  
Note that $N \sum_i p_i^2$ is related to the second and first moments of the degree distribution $P(k)$,
$\mediaT{k}_{_P}=c$ and $\mediaT{k^2}_{_P}$,
in terms of which Eq. (\ref{THEOcritpure}) becomes identical to the critical surface valid
for the pure scale free graph obtained by using the configuration model \cite{GZ} when
$\mediaT{k^2}_{_P}<\infty$ (note however that Eq. (\ref{THEOcritpure1}) 
is valid in general also when $1<2\mu<2$ where $\mediaT{k^2}_{_P}=\infty$):
\begin{eqnarray}
\label{THEOcritpure2}
t_c^{(\Sigma)}\frac{\mediaT{k^2}_{_P}-\mediaT{k}_{_P}}{\mediaT{k}_{_P}}=1.
\end{eqnarray}
As anticipated before,
we find instead a complete novel result for the correlation function $\tilde{\chi}_{i,j}^{(\Sigma)}$ 
of two given spins $i$ and $j$. Let us consider only the case $\Sigma=$F in the P region
and let us consider the choice (\ref{stat})-(\ref{stat1}).
From Eq. (\ref{THEOcritpure1}) we see that, when $N$ grows, $t_c^{(\mathrm{F})}$ and $t^{(\mathrm{F})}$ in the P region
remain finite for $2\mu<1$, while they go to 0 for $2\mu>1$ (logarithmically for $2\mu=1$),
therefore, in the latter case, for finite $N$, we can evaluate the correlation
function at a temperature scaling with the critical one.
In conclusion, from Eq. (\ref{THEOcritpure1}) and Eq. (\ref{THEOC2b}) 
applied with $\tilde{\chi}_{0;i,j}|_{\{h_l=0\}}=\delta_{i,j}$,
in the finite network,
for $2\mu<1$ and at any temperature above the critical one we have
\begin{eqnarray}
\label{THEOC2b0}
\tilde{\chi}_{i,j}^{(\mathrm{F})}=
\delta_{i,j}+\frac{t^{(\mathrm{F})}}{t_c^{(\mathrm{F})}}\frac{(1-2\mu)}
{(1-t^{(\mathrm{F})}/t_c^{(\mathrm{F})})}\frac{(ij)^{-\mu}}{N^{1-2\mu}},
\end{eqnarray}
whereas for $2\mu>1$ at any temperature scaling with the critical one and in the P region ($\beta<\beta_c$) we have 
\begin{eqnarray}
\label{THEOC2b01}
\tilde{\chi}_{i,j}^{(\mathrm{F})}=
\delta_{i,j}+\frac{\beta}{\beta_c^{(\mathrm{F})}}\frac{(2\mu-1)}
{(1-\beta/\beta_c^{(\mathrm{F})})}(ij)^{-\mu},
\end{eqnarray}
where we have made use of the fact that, up to negligible terms for $N$ large, 
$t^{(\mathrm{F})}/t_c^{(\mathrm{F})}=\beta/\beta_c^{(\mathrm{F})}$.
By using $p_i\simeq \media{k_i}/\sum_j\media{k_j}$, $\media{k_i}$ being the average degree of the vertex $i$,
we can express approximately Eqs. (\ref{THEOC2b0}) and (\ref{THEOC2b01}) in terms of the vertex degree as
\begin{eqnarray}
\label{THEOC2b0d}
\tilde{\chi}_{i,j}^{(\mathrm{F})}\simeq
\delta_{i,j}+\frac{t^{(\mathrm{F})}}{t_c^{(\mathrm{F})}}\frac{(1-2\mu)}
{(1-\mu)^2(1-t^{(\mathrm{F})}/t_c^{(\mathrm{F})})}\frac{\media{k_i}\media{k_j}}{\media{k}_{_P}^2N}, 
\end{eqnarray}
for $2\mu<1$, and
\begin{eqnarray}
\label{THEOC2b01d}
\tilde{\chi}_{i,j}^{(\mathrm{F})}\simeq
\delta_{i,j}+\frac{\beta}{\beta_c^{(\mathrm{F})}}\frac{(2\mu-1)}
{(1-\mu)^2(1-\beta/\beta_c^{(\mathrm{F})})}\frac{\media{k_i}\media{k_j}}{\media{k}_{_P}^2N^{2\mu}},
\end{eqnarray}
for $2\mu>1$.
However, by using only Eqs. (\ref{THEOC2b}) and (\ref{THEOcritpure}),
we can get the correlation function in a form which is completely independent of the form for the $p_i$'s
\begin{eqnarray}
\label{THEOC2b0e}
\tilde{\chi}_{i,j}^{(\mathrm{F})}\simeq
\delta_{i,j}+
\frac{t^{(\mathrm{F})}}{(1-t^{(\mathrm{F})}/t_c^{(\mathrm{F})})}
\frac{\media{k_i}\media{k_j}}{\media{k}_{_P}N}. 
\end{eqnarray}
Comparison of Eq. (\ref{THEOC2b0e}) with Eqs. (\ref{THEOC2b0}) and (\ref{THEOC2b01})
shows that the strongest correlations involve the nodes $i$'s with the highest degree with:
$\media{k_i}\sim \mathop{O}(N^{\mu})$ for $\mu<1/2$ ($\gamma>3$), and 
$\media{k_i}\sim \mathop{O}(N^{1/2})$ for $\mu\geq 1/2$ ($\gamma\leq 3$). 
 
\subsection{``Gas'' of dimers in a scale-free network}
Here we consider the case in which $(\mathcal{L}_0,\Gamma_0)$ is a set
of $N$ disconnected dimers (so that there are $2N$ sites).
This case represents the simplest example with $J_0\neq 0$ in which 
$m_{0i}(\beta J_0^{(\Sigma)};\{\beta h_j\})$ can be exactly calculated.
We have
\begin{eqnarray}
\label{THEOdim}
m_{0i}(\beta J_0^{(\Sigma)};\{\beta h_j\})=\frac{\tanh(\beta h_i)+t_0^{(\Sigma)}\tanh(\beta h_{j_0(i)})}
{1+t_0\tanh(\beta h_i)\tanh(\beta h_{j_0(i)})},
\end{eqnarray} 
where $t_0^{(\Sigma)}=\tanh(\beta J_0^{(\Sigma)})$ and $j_0(i)$ stands for the first neighbor of $i$.
By derivation we get the correlation function of the pure model 
$\tilde{\chi}_{0;i,j}$ which, in the P region, takes the form
\begin{eqnarray}
\label{THEOdim1}
\tilde{\chi}_{0;i,j}(\beta J_0^{(\Sigma)};0)=\left\{
\begin{array}{l}
1, \quad j=i \\
t_0^{(\Sigma)}, \quad j=j_0(i).
\end{array}
\right.
\end{eqnarray} 
Therefore for the critical surface we have
\begin{eqnarray}
\label{THEOdim2}
ct_c^{(\Sigma)}N \left[\sum_i p_i^2+t_0^{(\Sigma)}\sum_i p_ip_{j_0(i)}\right]=1.
\end{eqnarray}
With respect to the critical surface of the model with $J_0=0$ (the above Viana-Bray case) 
we see in Eq. (\ref{THEOdim2}) the presence of a term proportional to $t_0^{(\Sigma)}$.
How much this term affects $t_c^{(\Sigma)}$ depends on how the dimers are placed,
\textit{i.e.}, on how we choose the first neighbors $\{j_0(i)\}$.
Since by definition the dimers are not connected, in general for $j_0(i)$ we can
take $j_0(i)=i+k, \mod N$ where $k$ is a constant integer in the range $[1,N]$.
The exact evaluation of $t_c^{(\Sigma)}$ for $N$ large remains simple only if
$k$ does not depend on $N$ or $k=\mathop{O}(N)$. 
Under the choice (\ref{stat})-(\ref{stat1}), for the former case we get
\begin{eqnarray}
\label{THEOdim3}
ct_c^{(\Sigma)}\frac{(1+t_{c0}^{(\Sigma)})(1-\mu)^2}{(1-2\mu)}(1-N^{2\mu-1})=1,
\end{eqnarray}     
whereas for the latter the critical surface remains not affected by
$t_0^{(\Sigma)}$ as in Eq. (\ref{THEOcritpure1}).
When $k$ does not grow with $N$, for the correlation function for the $\Sigma=$F case in the P region 
for $2\mu<1$ we have
\begin{eqnarray}
\label{THEOdim4}
\tilde{\chi}_{i,j}^{(\mathrm{F})}&=&
\tilde{\chi}_{0;i,j}+\frac{t^{(\mathrm{F})}}{t_c^{(\mathrm{F})}}\frac{(1-2\mu)}
{(1+t_0^{(\mathrm{F})})(1-t^{(\mathrm{F})}/t_c^{(\mathrm{F})})}\times\nonumber \\
&& \frac{[i^{-\mu}+t_0^{(\mathrm{F})}(j_0(i))^{-\mu}][j^{-\mu}+t_0^{(\mathrm{F})}(j_0(j))^{-\mu}]}{N^{1-2\mu}},
\end{eqnarray}
whereas for $2\mu>1$ at a temperature scaling with the critical one we have 
\begin{eqnarray}
\label{THEOC2b01V}
\tilde{\chi}_{i,j}^{(\mathrm{F})}&=&
\tilde{\chi}_{0;i,j}+\frac{\beta}{\beta_c^{(\mathrm{F})}}\frac{(2\mu-1)}
{(1+t_0^{(\mathrm{F})})(1-\beta/\beta_c^{(\mathrm{F})})}\times\nonumber \\
&& [i^{-\mu}+t_0^{(\mathrm{F})}(j_0(i))^{-\mu}][j^{-\mu}+t_0^{(\mathrm{F})}(j_0(j))^{-\mu}].
\end{eqnarray} 
Similar expressions hold for the correlation in the case in which $k=\mathop{O}(N)$, the
only difference being the absence of the pref-actor $1/(1+t_0^{(\mathrm{F})})$. 
More in general, independently of the form for the $p_i$'s we, in terms of the average degrees we have
\begin{eqnarray}
\label{THEOC2b01V1}
&&\tilde{\chi}_{i,j}^{(\mathrm{F})}\simeq
\delta_{i,j}+\frac{t^{(\mathrm{F})}}{(1-t^{(\mathrm{F})}/t_c^{(\mathrm{F})})} \nonumber \\ && \times
\frac{\left[\media{k_i}+t_0^{(\mathrm{F})}\media{k_{j_0(i)}}\right]
\left[\media{k_j}+t_0^{(\mathrm{F})}\media{k_{j_0(j)}}\right]}{\media{k}_{_P}N}. 
\end{eqnarray}

\subsection{A one dimensional chain through the scale-free network}
Here we consider the case in which $(\mathcal{L}_0,\Gamma_0)$ is a 
a one dimensional chain with periodic boundary conditions
and such that the first site of the chain corresponds to the site $i=1$ of the static network,
the second site of the chain corresponds to the site $i=2$ of the static network, and so on.
As we have learned in Sec. III C, this or any other choice will not alter the critical behavior
of the whole system that remains the same as in the absence of the chain.
In the P region the correlation function of the pure model is given by
\begin{eqnarray}
\label{THEO1d}
\tilde{\chi}_{0;i,j}(\beta J_0^{(\Sigma)};0)=[t_0^{(\Sigma)}]^{|j-i|},
\end{eqnarray} 
from which by using Eq. (\ref{THEOcrit}) we get the critical surface
\begin{eqnarray}
\label{THEO1d1}
ct_c^{(\Sigma)}N \sum_{i,j}p_ip_j [t_0^{(\Sigma)}]^{|j-i|}=1.
\end{eqnarray}
Let us consider the choice (\ref{stat})-(\ref{stat1}).
For $|t_0^{(\Sigma)}|<<1$ we need to keep track only of the term $\mathop{O}(t_0^{(\Sigma)})$ and for $N$ large we get
\begin{eqnarray}
\label{THEO1d2}
ct_c^{(\Sigma)}\frac{(1+2t_{c0}^{(\Sigma)})(1-\mu)^2}{(1-2\mu)}(1-N^{2\mu-1})=1.
\end{eqnarray}     
In general Eq. (\ref{THEO1d2}) is exact only in the region $2\mu\geq 1$ so that $t_{c0}^{(\Sigma)}\to 0$ for $N\to\infty$.
Notice the difference with respect to the gas of dimers case
in Eq. (\ref{THEOdim3}) for the presence of a factor 2 in front of the term proportional to $t_{c0}^{(\Sigma)}$.
When $|t_0^{(\Sigma)}|<<1$, for the correlation function for the $\Sigma=$F case in the P region 
with $2\mu<1$ we have
\begin{eqnarray}
\label{THEO1d3}
\tilde{\chi}_{i,j}^{(\mathrm{F})}&\simeq&
[t_0^{(\mathrm{F})}]^{|j-i|}+\frac{t^{(\mathrm{F})}}{t_c^{(\mathrm{F})}}\frac{(1-2\mu)}
{(1-t^{(\mathrm{F})}/t_c^{(\mathrm{F})})}\times\nonumber \\
&& \frac{(1+2t_0^{(\mathrm{F})})(ij)^{-\mu}}{N^{1-2\mu}},
\end{eqnarray}
whereas for $2\mu>1$ at a temperature scaling with the critical one we have (with a better approximation)
\begin{eqnarray}
\label{THEO1d4}
\tilde{\chi}_{i,j}^{(\mathrm{F})}&\simeq&
[t_0^{(\mathrm{F})}]^{|j-i|}+\frac{\beta}{\beta_c^{(\mathrm{F})}}\frac{(2\mu-1)}
{(1-\beta/\beta_c^{(\mathrm{F})})}\times\nonumber \\
&& (1+2t_0^{(\mathrm{F})})(ij)^{-\mu},
\end{eqnarray} 
where we have approximated $i+1\simeq i-1 \simeq i$.
The analytical evaluation of the lhs of Eq. (\ref{THEO1d1}) for $t_0^{(\Sigma)}$ finite remains a difficult task
and we have to resort to a numerical evaluation at a sufficiently large value of $N$ such that
finite size effects become negligible. As we have just learned, finite size
effects can have a very slow relaxation rate in scale free graphs;
in evaluating the correlation functions, when $\mu<1/2$ ($\gamma> 3$)
they decay as slowly as $1/N^{1-2\mu}$, while  
they persist even in the thermodynamic limit when $\mu>1/2$ ($\gamma\leq 3$).
As we have seen above, however, we can easily handle the latter case since $t_0^{(\Sigma)}$ is always small.
We see then that the most difficult numerical task
in the evaluation of the lhs of Eq. (\ref{THEO1d1}),
as well as in general formulas involving sums of correlation functions, 
occurs in the case of a distribution
with $\mu\to 1/2^{-}$ ($\gamma \to 3^{-}$). 

\begin{figure}[tbh]
\epsfxsize=90mm  \centerline{\epsffile{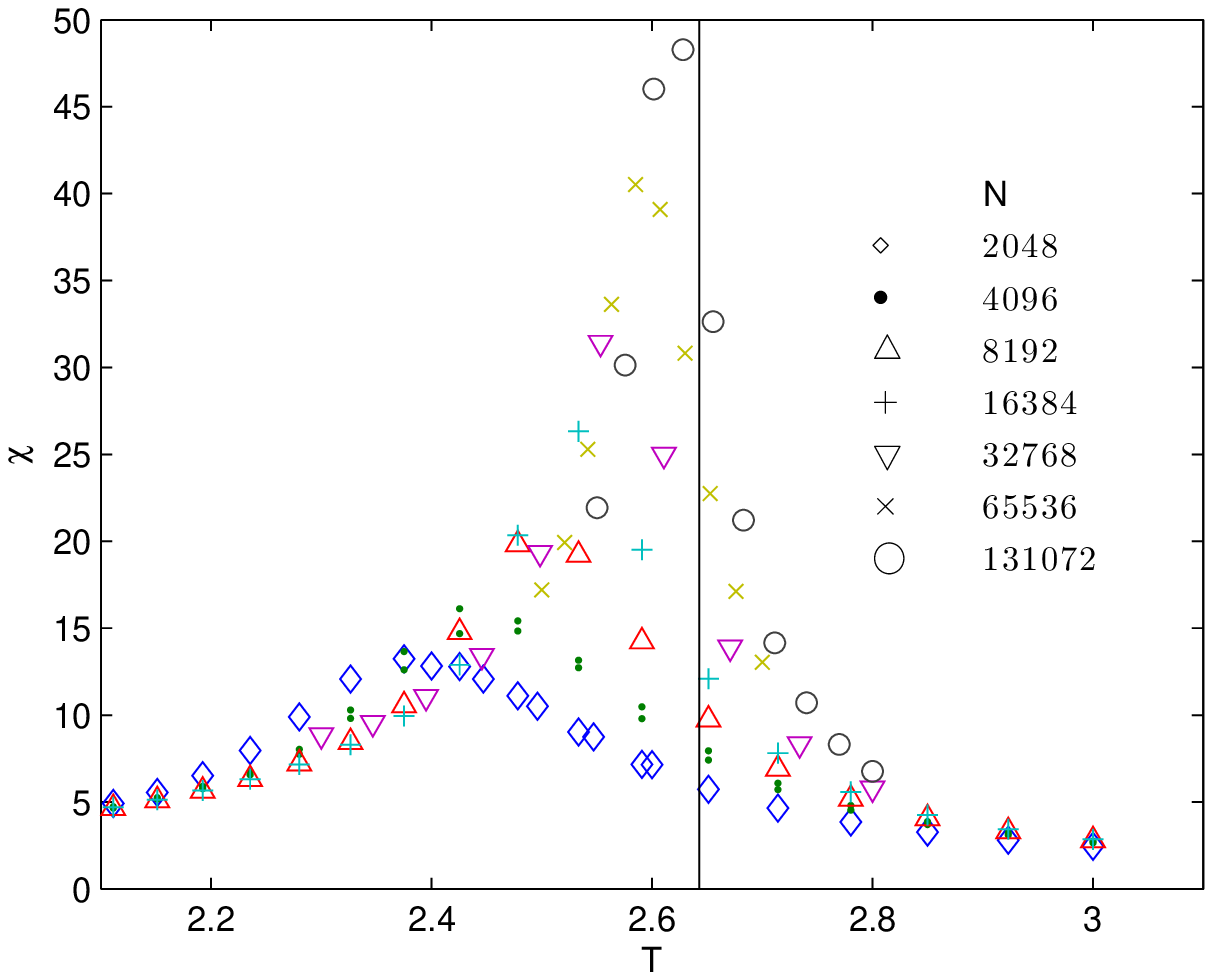}}
\centerline{\epsffile{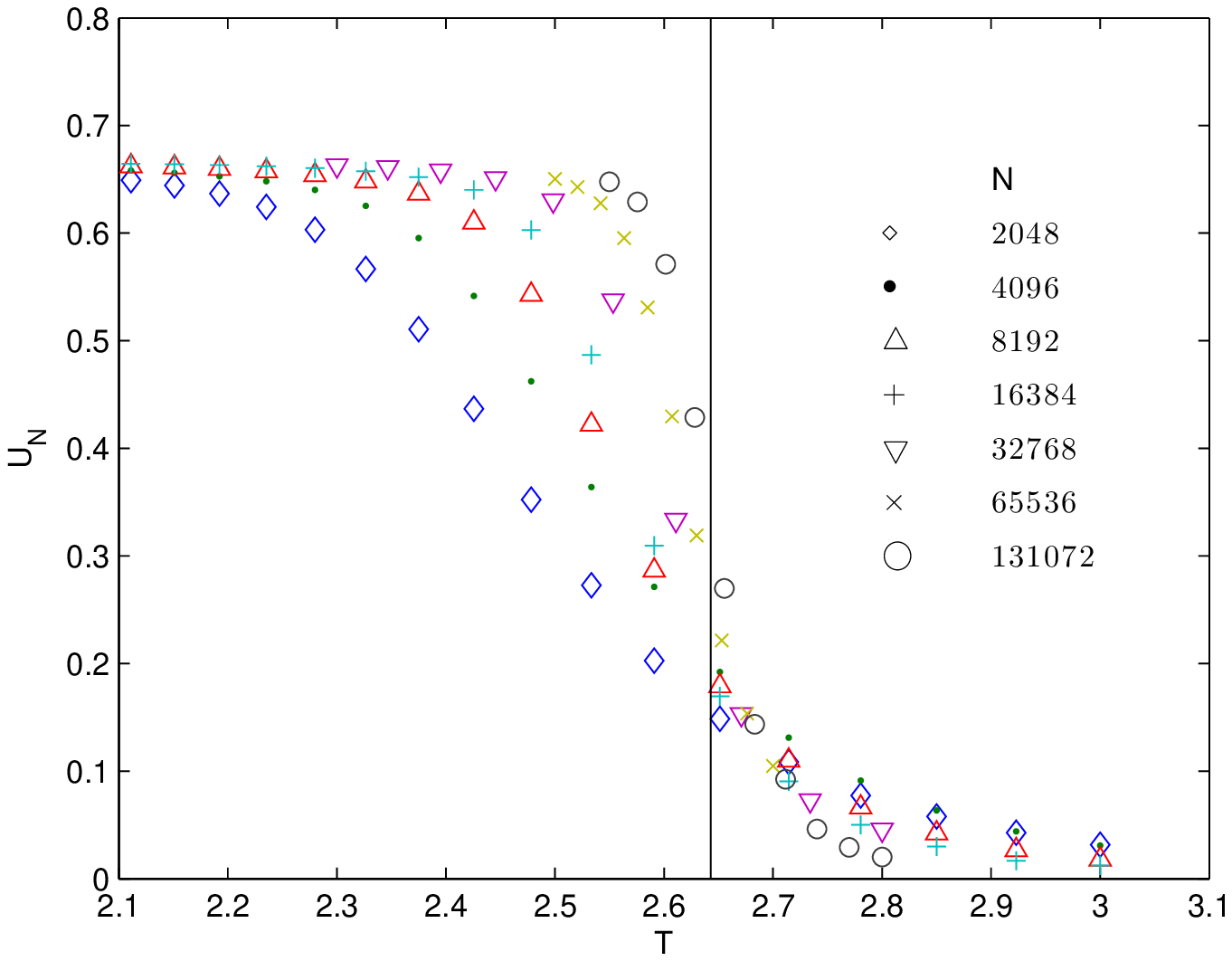}}
\caption{\label{fig1}(Color online) Plots of the susceptibility $\chi$ (top panel), 
and of the Binder Cumulant $U$ (bottom panel), as a function of the temperature $T$ for the 
random model (\ref{H})-(\ref{hidden}) in which $(\mathcal{L}_0,\Gamma_0)$ 
is a one dimensional chain and the random network is
generated via the choice (\ref{stat})-(\ref{stat1}) with $\mu=1/3$ (corresponding to $\gamma=4$).
The other parameters of the model are: $c=1$, $J=J_0=1$, and $p=0$. 
The vertical line comes from the solution of Eq. (\ref{THEO1d1}) with $N=131072$.}
\end{figure}

\begin{figure}[tbh]
\epsfxsize=90mm  \centerline{\epsffile{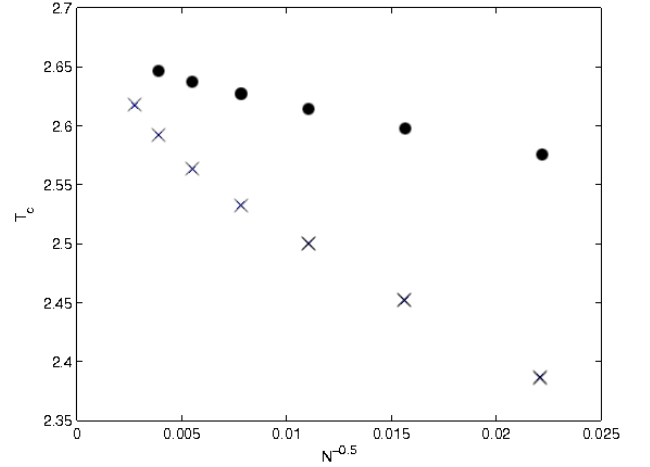}}
\caption{\label{fig3} 
Plots of the ``finite size critical temperature'' $T_c(N)$ as a function of the system size $N$
for theoretical (dots) and for simulation data (crosses) of the same system of Fig. (\ref{fig1}).
We stress that, though the accordance between the theoretical and the simulation data is poor for 
relatively small values of $N$, as explained at the end of Sec. IVC,
the two trends for large $N$ must fit with a $N^{1/2}$ behavior and approach the same
value in the limit $N\to\infty$. The figure confirm our analysis.}
\end{figure}

In Figs. (\ref{fig1}) we plot simulations for the susceptibility $\chi$ 
and for the Binder Cumulant $U$ \cite{Binder}, respectively, as a function of the temperature 
$T$ for several system sizes $N$ and compare the location of the maximums with the theoretical $T_c$
evaluated at a very large value of $N$ where we observe stationarity within the statistical errors. 
Finally in Fig. (\ref{fig3}), for growing but finite sizes $N$, 
we plot the position of the ``finite size $T_c(N)$'', defined as
the position of the maximum of the susceptibility $\chi$ with respect to the temperature.
We evaluate such quantities for both simulations and theoretical data of the same system as a function of $N^{1/2}$.
Note that the latter evaluation coincides simply with the solution coming from Eq.(\ref{THEO1d1}).   
From Fig. (\ref{fig3}) we find confirmation of two facts: 
\textit{i)} Eq.(\ref{THEO1d1}) (as well as all the effective field theory in general) 
has a clear meaning also at finite sizes;
\textit{ii)} since for a mean-field universality class it is expected to be at criticality 
$\tilde{\chi}_c(N)\sim N^{1/2}$ \cite{Luijten1995}, as also confirmed in \cite{SWMC},
and since, on the other hand, from Eq. (\ref{THEOchie2}) for finite $N$ 
we have $\tilde{\chi}_c(N)\sim \mathop{O}(1)/(T_c(N)-T_c)$,
we get $T_c(N)\sim T_c+\mathop{O}(1)/{N}^{1/2}$, in accordance with Fig. (\ref{fig3}).


\section{Application to network design}
\label{design}
As we have seen in Sec. IIIB2, 
if $(\mathcal{L}_0,\Gamma_0)$ is a homogeneous graph,
\textit{i.e.}, its adjacency matrix $c_{0;i,j}$ has some periodicity, 
clustering increases the percolation threshold.
A different question arises instead if between the
$\{p_i\}$ and the $\{c_{0;i,j}\}$ there is some correlation.
Given the desired degree sequence and then the weights $\{p_i\}$,
we see that, if $(\mathcal{L}_0,\Gamma_0)$ is not translational invariant, 
we can optimize the percolation by labeling the sites
in such a way that the functional $F_\infty(\{p_i\})$ is maximized, where  
(as before here $p$ is the probability by which each link of the pure graph $(\mathcal{L}_0,\Gamma_0)$ is removed)
\begin{eqnarray}
\label{THEOcrit6}
F_\infty(\{p_i\})\defi cN\sum_{i,j}\tilde{\chi}_{0;i,j}(\tanh^{-1}(1-p);0)p_ip_j.
\end{eqnarray} 
Alternatively, Eq. (\ref{THEOcrit6}) can be rewritten 
in terms of the graph elements of $(\mathcal{L}_0,\Gamma_0)$ as
\begin{eqnarray}
\label{THEOcrit6b}
F_\infty(\{p_i\})= cN \sum_{i,j}\left(\delta_{i,j}+\mathcal{N}_{0;i,j}^{(p)}\right)p_ip_j.
\end{eqnarray}
where $\mathcal{N}_{0;i,j}^{(p)}=1$ if, in the graph $(\mathcal{L}_0,\Gamma_0)$
from which each link has been removed at random with probability $p$,
between the vertex $i$ and the vertex $j$ there exists at least one path of links, 
and $\mathcal{N}_{0;i,j}^{(p)}=0$ otherwise.

Once the $\{p_i\}$ that optimizes $F_\infty(\{p_i\})$ has been found
through a suitable labeling,
the corresponding network will have in general a percolation threshold $c_c$ given by the equation $F_\infty(\{p_i\})=1$
that is a minima with 
respect to all the possible $N!$ labelings. 
From Eq. (\ref{THEOcrit6}) we see that a simple approximate heuristic 
receipt to approach this optimum consists in choosing a labeling of the weights $\{p_i\}$,
$i_1,i_2,\ldots,i_N$, such that 
\begin{eqnarray}
\label{Heuristic}
p_{i}\geq p_{j},\quad \Leftrightarrow \quad \tilde{\chi}_{0;i}\geq \tilde{\chi}_{0;j}, \quad \forall i,j\in \mathcal{L}_0
\end{eqnarray}
where $\tilde{\chi}_{0;i}(\tanh^{-1}(1-p);0)$
stands for the total correlation of the graph $(\mathcal{L}_0,\Gamma_0)$ at zero temperature at the point $i$:
\begin{eqnarray}
\label{THEOcrit7}
\tilde{\chi}_{0;i}\defi \sum_j \tilde{\chi}_{0;i,j}(\tanh^{-1}(1-p);0),
\end{eqnarray} 
or alternatively
\begin{eqnarray}
\label{THEOcrit7b}
\tilde{\chi}_{0;i}=1+ \sum_j \mathcal{N}_{0;i,j}^{(p)},
\end{eqnarray} 
\textit{i.e.}, $\tilde{\chi}_{0;i}$ can be seen as 
the average total number of nodes connected to $i$ (including the node $i$ itself)
in the graph $(\mathcal{L}_0,\Gamma_0)$
from which each link has been removed at random with probability $p$.

Our optimal design problem can be precisely formulated as follows.
Given a graph $(\mathcal{L}_0,\Gamma_0)$, in which each link is removed with
probability $p$, and given a desired additional degree
sequence $\{\bar{k}_i\}$ (proportional to the weights $\{p_i\}$) having an average connectivity $c$,
we have to place the $L=cN/2$ additional ``long-range'' links on $(\mathcal{L}_0,\Gamma_0)$ in such a way 
that the resulting graph has a maximal percolating cluster.
Within our effective field theory this task amounts to say that $F_\infty(\{p_i\})$ is maximized.
As we have explained in \cite{CommuL}, however, a network, at least in the absence
of congestion \cite{Guimera}, benefits 
from optimal communication features at the percolation point. Adding further links
after this point makes the network less sensitive to signals.
On the other hand if, for a given value of $c$, we have found an optimal labeling which
maximizes $F_\infty(\{p_i\})$, from Eq. (\ref{THEOcrit6}) 
we see that changing only $c$ will leave 
still the choice of the

labeling as an extremal choice for $F_\infty(\{p_i\})$.
Therefore, we can speak of label optimization for $F_\infty(\{p_i\})$ regardless
of the value of $c$. In particular, after finding the optimal labeling, we will be free
to choose for $c$ a value such that $F_\infty(\{p_i\})=1$, so that we will be in the percolation threshold.
Among all the possible other labelings the network will have a minimal percolation threshold. 
This implies that for the found optimal labeling 
the graph will benefit of optimal communication features but with minimal cost (if the cost is given by $c$).
Our design strategy is therefore in the same philosophy of \cite{Munoz}
where the optimality was defined with respect to synchronization.
Although in general these two different criteria of design may give rise
to different networks, we argue that, in the absence of congestion, some general properties about
efficient communication are shared~\footnote{
We think in that if our optimal-percolation criterion is equipped 
with further constrains on the $\{p_i\}$ the two keys of design would share more and more properties}.
 
A particularly interesting case is the one in which the graph $(\mathcal{L}_0,\Gamma_0)$
is splitted into disjoint subsets, that we could then call isolated communities.
In \cite{CommuL} we had considered the problem of percolation for a generic set of $n$ communities,
isolated or not, in which the additional ``long-range'' links were defined through an additional
$n\times n$ matrix $\bm{c}$ of intra- (inside the community) and inter- (among the communities) connectivities. 
In that problem the unknown quantity
to be found was the critical matrix $\bm{c}$ at which percolation sets in, 
and it was easy to find that the critical $\bm{c}$ must satisfy the equation
\begin{eqnarray}
\label{Commu}
\det(\bm{1}-\tilde{\bm{\chi_0}}\cdot\bm{c})=0,   
\end{eqnarray}
$\tilde{\bm{\chi_0}}$ being the matrix of the relative intra- and inter-susceptibilities among the communities.
In particular Eq. (\ref{Commu}) for the case in which $(\mathcal{L}_0,\Gamma_0)$ is simply a disjoint set of nodes becomes
\begin{eqnarray}
\label{Commu1}
\det(\bm{c})=1,   
\end{eqnarray} 
which constitutes a clean generalization of the well known percolation threshold $c=1$ of the
case $n=1$ corresponding to the Erd$\mathrm{\ddot{o}}$s-R$\mathrm{\acute{e}}$nyi random graph \cite{Classical}.
However, Eq. (\ref{Commu}) is a single equation in the $n\times n$ unknown matrix elements of $\bm{c}$, $c^{(l,m)}$,
therefore there are infinite solutions for $n>1$. 
Given the matrix $\tilde{\bm{\chi_0}}$, now,
the analogous optimization problem that we have defined before amounts to look for the
matrix $\bm{c}$ that gives the maximum value of the largest eigenvalue of the matrix $\tilde{\bm{\chi_0}}\cdot\bm{c}$
under the constraint that the total cost is fixed: $\sum_{l,m}c^{(l,m)}/n=c$ \cite{Commu}.
In general, in this kind of problem one can find several solutions that represent local maximum,
but the asymmetric ones, if any, are those which guarantee better communication performance \cite{CommuL}.
This fact is reminiscent of the so called ``star-like'' configuration,
which is known to provide the best communication/searchability performance 
in the absence of congestion \cite{Guimera}.    
In the problem considered in \cite{CommuL}, however, the additional ``long-range'' links were uniformly
distributed among the communities, though with the use of $n\times n$ different average 
connectivities $c^{(l,m)}$.
Here we face instead the problem in which we have one single additional connectivity $c$, but
the ``long-range'' links follow a generic desired degree sequence $\{p_i\}$, which in particular can be scale free. 
We observe that, in the case in which the graph $(\mathcal{L}_0,\Gamma_0)$
is splitted into $n$ disjoint communities, $(\mathcal{L}_0^{(l)},\Gamma_0^{(l)})$, $l=1,\ldots,n$, 
the heuristic solution (\ref{Heuristic}) 
provides the exact minima for $F_\infty(\{p_i\})$ 
when all the communities have equal size and 
are internally homogeneous (\textit{i.e.}, $\tilde{\chi}_{0;i,j}$ is the same
for any $i,j\in \mathcal{L}_0^{(l)}$). 

Let us consider, for example, the case in which we have $n$ disjoint communities of size $N^{(l)}$, $l=1,\ldots,N$
such that $N=\sum_l N^{(l)}$. Let us suppose that each community $(\mathcal{L}_0^{(l)},\Gamma_0^{(l)})$
consists in a homogeneous random graph with average connectivity $c_0^{(l)}<1$
(we want to consider only situations in which the set
$(\mathcal{L}_0,\Gamma_0)$ is not already a percolating cluster). 
From Eq. (\ref{THEOC2d}) we have simply
\begin{eqnarray}
\label{Commu2}
\tilde{\chi}_{0;i,j}=\delta_{i,j}+\frac{c_0^{(l)}}{N^{(l)}\left[1-c_0^{(l)}\right]}, ~i,j\in \mathcal{L}_0^{(l)},~  
l=1,\ldots,n,\nonumber \\
\end{eqnarray} 
from which we get the following total correlator (Eq. (\ref{THEOcrit7}) with the choice $p=0$) which depends only on 
the community index $l$:
\begin{eqnarray}
\label{Commu3}
\tilde{\chi}_{0}^{(l)}\defi \tilde{\chi}_{0;i}=1+\frac{c_0^{(l)}}{1-c_0^{(l)}}, \quad i\in \mathcal{L}_0^{(l)},  
\quad l=1,\ldots,n.\nonumber \\
\end{eqnarray} 
Taking into account that the $n$ communities of $(\mathcal{L}_0,\Gamma_0)$ do not interact 
(there is no inter-link) by plugging Eq. (\ref{Commu2}) in Eq. (\ref{THEOcrit6}) we have 
\begin{eqnarray}
\label{Commu4}
&&F_\infty(\{p_i\})=cN\sum_i p_i^2 \\
&&+ \sum_{l=1}^n \frac{cc_0^{(l)}}{\alpha^{(l)}\left[1-c_0^{(l)}\right]}
\left[\left(\sum_{i\in \mathcal{L}_0^{(l)}}p_i\right)^2-\sum_{i\in \mathcal{L}_0^{(l)}}p_i^2\right],\nonumber
\end{eqnarray} 
where $\alpha^{(l)}\defi N^{(l)}/N$.
For $N$ large Eq. (\ref{Commu4}) becomes
\begin{eqnarray}
\label{Commu5}
&&F_\infty(\{p_i\})=cN\sum_i p_i^2
\nonumber \\ &&
+ c\sum_{l=1}^n \frac{c_0^{(l)}}{\alpha^{(l)}\left[1-c_0^{(l)}\right]}
\left(\sum_{i\in \mathcal{L}_0^{(l)}}p_i\right)^2,
\end{eqnarray}
which, in turn, can be rewritten in terms of the $\tilde{\chi}_{0}^{(l)}$ as
\begin{eqnarray}
\label{Commu6}
&&F_\infty(\{p_i\})=cN\sum_i p_i^2
\nonumber \\ &&
+ c\sum_{l=1}^n \frac{\tilde{\chi}_{0}^{(l)}-1}{\alpha^{(l)}}
\left(\sum_{i\in \mathcal{L}_0^{(l)}}p_i\right)^2,
\end{eqnarray} 
From Eq. (\ref{Commu6}) we see that the heuristic solution (\ref{Heuristic})
provides manifestly the global maximum for $F_\infty(\{p_i\})$ only when
all the communities have the same size $\alpha^{(l)}=1/n$, otherwise the exact global
maximum will be provided by the labelings of the $p_i$'s such that
\begin{eqnarray}
\label{HeuristicCommu}
&& p_{i}\geq p_{j},\quad \Leftrightarrow \quad \frac{\tilde{\chi}_{0}^{(l)}-1}{\alpha^{(l)}}\geq 
\frac{\tilde{\chi}_{0}^{(m)}-1}{\alpha^{(m)}}, 
\nonumber \\ &&
\quad \forall i \in \mathcal{L}_0^{(l)},\forall j \in \mathcal{L}_0^{(m)}.
\end{eqnarray}
Of course, due to the homogeneity of the communities, we have at least as many equivalent global maximum as 
$\prod_{l=1}^nN^{(l)}!$ (if the ratios $[\tilde{\chi}_{0}^{(l)}-1]/\alpha^{(l)}$ are not all different
the number of equivalent global maximum is greater).

In the above example we had three important simplifications: 
\textit{(i)} the communities were not interacting; 
\textit{(ii)} each community was homogeneous; 
\textit{(iii)} we were able to calculate analytically the terms $\tilde{\chi}_{0;i,i}$.
In the most general case none of the above conditions is satisfied.
In particular, when condition \textit{(iii)} is not satisfied, to calculate
the total correlator $\tilde{\chi}_{0}^{(l)}$ of the $l$-th community, defined as 
\begin{eqnarray}
\label{CorreCommu}
\tilde{\chi}_{0}^{(l)}\defi \sum_{i,j\in\mathcal{L}_0^{(l)}}\tilde{\chi}_{0;i,j},
\quad l=1,\ldots,n,
\end{eqnarray} 
we have to resort to a Monte Carlo strategy, either by an Ising model at low temperature (simulated annealing) 
(in view of Eq. (\ref{THEOcrit6})) or by a direct graph analysis (in view of Eq. (\ref{THEOcrit6b})).
However, if conditions \textit{(i)} and \textit{(ii)} are satisfied, Eq. (\ref{Commu6}) 
is still exact and the global maximum will be given by Eq. (\ref{HeuristicCommu}) with the total
correlator defined by Eq. (\ref{CorreCommu}).
It is interesting to note that if, as usually happens, the communities are 
hierarchically organized in nested communities at deeper and deeper levels, 
and conditions \textit{(i)} and \textit{(ii)} still satisfied at each level of the hierarchy, we can iterate
the above procedure through a natural generalization of Eqs. (\ref{Commu6})-(\ref{CorreCommu}) 
at each level of the hierarchy.
For example, if the communities are organized into two levels, \textit{i.e.},
$(\mathcal{L}_0,\Gamma_0)$ is splitted in $n_1$ communities $l_1=1,\ldots,n_1$, each one in turn splitted in $n_{l_1}$
communities as
\begin{eqnarray}
\label{Hierachy}
(\mathcal{L}_0,\Gamma_0)&&=\cup_{l_1=1}^{n_1}(\mathcal{L}_0^{(l_1)},\Gamma_0^{(l_1)})
\nonumber \\ &&
=\cup_{l_1=1}^{n_1}\cup_{l_2=1}^{n_{l_1}} (\mathcal{L}_0^{(l_1,l_2)},\Gamma_0^{(l_1,l_2)}),
\end{eqnarray}
it is then easy to see that Eq. (\ref{Commu6}) generalizes to
\begin{eqnarray}
\label{HCommu6}
&&F_\infty(\{p_i\})=cN\sum_i p_i^2
\nonumber \\ &&
+ c\sum_{l_1=1}^{n_1}\sum_{l_2=1}^{n_{l_1}} \frac{\tilde{\chi}_{0}^{(l_1,l_2)}-1}{\alpha^{(l_1,l_2)}}
\left(\sum_{i\in \mathcal{L}_0^{(l_1,l_2)}}p_i\right)^2,
\end{eqnarray} 
which has a global maximum in correspondence of the following labeling, natural
generalization of Eq. (\ref{HeuristicCommu}):
\begin{eqnarray}
\label{HH}
&& p_{i}\geq p_{j},\quad \Leftrightarrow \quad \frac{\tilde{\chi}_{0}^{(l_1,l_2)}-1}{\alpha^{(l_1,l_2)}}\geq 
\frac{\tilde{\chi}_{0}^{(m_1,m_2)}-1}{\alpha^{(m_1,m_2)}}, 
\nonumber \\ &&
\quad \forall i \in \mathcal{L}_0^{(l_1,l_2)},\forall j \in \mathcal{L}_0^{(m_1,m_2)},
\end{eqnarray}
where now the total correlators $\tilde{\chi}_{0}^{(l_1,l_2)}$ and the coefficients $\alpha^{(l_1,l_2)}$ are defined as 
\begin{eqnarray}
\label{HCorreCommu}
\tilde{\chi}_{0}^{(l_1,l_2)}\defi \sum_{i,j\in\mathcal{L}_0^{(l_1,l_2)}}\tilde{\chi}_{0;i,j},
\end{eqnarray}  
\begin{eqnarray}
\label{HCorreCommu}
\alpha^{(l_1,l_2)}\defi \frac{N}{N^{(l_1,l_2)}},
\end{eqnarray}  
with $N^{(l_1,l_2)}\defi|\mathcal{L}_0^{(l_1,l_2)}|$.

Whatever be the graph $(\mathcal{L}_0,\Gamma_0)$,
the task to compute via a Monte Carlo method the total correlators 
usually requires a computational cost which
grows only polynomially in the system size $N$. 
A serious problem comes however when 
conditions \textit{(i)} or \textit{(ii)} are not satisfied. In this case in fact 
the heuristic solution (\ref{HeuristicCommu}) (or its generalization to the hierarchical case) in general
will not provide the global maximum for $F_\infty(\{p_i\})$.
As an intermediate situation it may happen that condition \textit{(i)} is not exactly satisfied, but 
the interaction among different communities
is weak so that the heuristic solution (\ref{HeuristicCommu}) (or its generalizations), via the evaluation
and comparison of the total correlators, is still a good
starting point for the numerical search of the exact global maximum of $F_\infty(\{p_i\})$,
especially when also condition \textit{(ii)} is almost satisfied. 
However, when the communities are well connected each other, or
there is no community structure at all, 
$F_\infty(\{p_i\})$ in general presents an exponential number of local maximum, and in fact
the computational complexity of the search for the global maximum of $F_\infty(\{p_i\})$
becomes equivalent to the Traveling Salesman Problem, which is an NP-hard problem \cite{TSP}.
In this case, in the graph $(\mathcal{L}_0,\Gamma_0)$ there is an high degree of frustration
and the heuristic ansatz (\ref{Heuristic}) might be very far,
not only from the exact global solution, but in general also from the local solutions.
We conclude however stressing that, despite this worst case scenario for the most general
optimization problem, in which one is forced to check for almost all the possible
$N!$ labelings of the $p_i$'s, the optimization of $F_\infty(\{p_i\})$ remains still
exponentially advantageous with respect to a direct inspection (in which $c$ is supposed to be given)
of all the possible graphs that one can build up by adding $L=cN/2$ long-range links over the
graph $(\mathcal{L}_0,\Gamma_0)$. 
In fact, given $c$, if we evaluate the number of ways $\mathcal{N}_c$ to lie $L=cN/2$ long-range links among $N$
nodes, for $c$ finite and $N$ large we get 
\begin{eqnarray}
\label{Cost}
&& \mathcal{N}_c=
\left(
\begin{array}{c}
\frac{N(N-1)}{2}\\
\\
\frac{cN}{2}
\end{array}
\right) \\ &&
\sim \exp\left[\frac{N(N-1)}{2}+\left(\frac{c}{2}-1\right)N\log(N)\right]\gg N!. \nonumber
\end{eqnarray}  

\section{Conclusions}
In this paper we have considered in detail, and in a more general
framework, the heterogeneous small-world model  
which was briefly presented in the Letter \cite{SWSF}, providing now all the complete proofs
and new applications. 
By using an effective field theory we prove in particular 
that the critical behavior is never affected by the presence of short loops (see Table I). 
We then apply the general result 
to the study of percolation, correlation functions and network design. 

By studying the percolation  we have shown, by considering several
analytically solvable examples, the role played by  short loops in
modifying the percolation threshold in networks.
In particular, we have seen how the presence of short loops 
increases the percolation point [see Eq. (\ref{THEOcritc5fe})]. 

By studying the correlation functions, we have found that for a
scale-free network, with or without short loops, finite size
effects can be very strong [see Table II and Eqs. (\ref{THEOh}) and
(\ref{delta})]. Moreover, when $\gamma$, the exponent of the
degree distribution, is as small as $\gamma\leq 3$, the finite
size effects become persistent even in the thermodynamic limit,
with the strongest correlations being those among hubs.
We stress that this is true even in the paramagnetic region
and with or without short-range couplings,
contradicting then the common opinion that correlations in purely
mean-field models always disappear in the thermodynamic limit
\footnote{This scenario is however compatible with the fact
that when $\gamma\leq 3$ a network can be ultra small-world 
with an average distance between nodes which can be of the order $\log(\log(N))$, 
or even finite in the thermodynamic limit \cite{UltraSmall}.}.

Finally, we have seen that the formula for the percolation
threshold suggests a natural way to optimize the communication
features among communities even if they interact.
We propose and discuss the efficiency of a heuristic solution 
[see Eqs. (\ref{Heuristic})-(\ref{THEOcrit7b})] at several levels:
isolated and homogeneous communities, weakly interacting communities, and
ill defined communities.   
The worst case scenario in which there is no evident community 
structure, is an NP-hard problem equivalent 
to the Traveling Salesman Problem, nevertheless, 
the use of the formula is still exponentially convenient with
respect to a direct inspection of the network.
We think that, at least in the absence of load-congestion, 
our algorithm can find important real-world applications.

\begin{acknowledgments}
This work was supported by
PTDC/FIS/108476/2008, PTDC/MAT/114515/2009 and SOCIALNETS.
We thank S. N. Dorogovtsev for useful discussions.
\end{acknowledgments}
\appendix

\section{Bounding $\mathcal{N}_N$}
By using, as in \cite{SatorrasHidden}, 
the approximation $\bar{k}_i/\sum_j \bar{k}_j\sim p_i$,
where $\bar{k}_i$ is the average degree of the vertex $i$, from Eq. (\ref{hidden3}) we have 
\begin{eqnarray}
\label{hidden3a}
\mathcal{N}_N= \sum_{i<j}\theta\left(\frac{\bar{k}_i\bar{k}_j}{Nc}-1\right),
\end{eqnarray}
which can be rewritten as
\begin{eqnarray}
\label{hidden3b}
\mathcal{N}_N= \frac{N}{2}\sum_i\mathcal{P}\left(k>\frac{Nc}{{k}_i}|k_i\right)p(k_i), 
\end{eqnarray}
where $p(k_i)$ is the probability that vertex $i$ has degree $k_i$, and 
$\mathcal{P}\left(k>\frac{Nc}{{k}_i}|k_i\right)$ is the conditional probability that, given that
the vertex $i$ has degree $k_i$, a randomly chosen vertex different from $i$ has degree greater than $Nc/k_i$.
Due to the weak degree-degree correlation of the network, 
from $\mathcal{P}\left(k>\frac{Nc}{{k}_i}|k_i\right)\simeq \mathcal{P}\left(k>\frac{Nc}{{k}_i}\right)$,
and from Eq. (\ref{hidden3b}) we have
\begin{eqnarray}
\label{hidden3c}
\mathcal{N}_N< \frac{N}{2}\sum_ip(k_i)\int_{\frac{Nc}{k_i}}^{k_M(N)} dk~ p(k),
\end{eqnarray}
where $k_M(N)$ is the maximum allowed degree in the network. Of course it always $k_M(N)\leq N$.
By using now the hypothesis that for $k$ large $p(k)\sim k^{-\gamma}$, we arrive at
\begin{eqnarray}
\label{hidden3c}
\mathcal{N}_N&<& \frac{N}{2(\gamma-1)}\int_1^{k_M(N)} dk~ p(k)k^{\gamma-1}(Nc)^{1-\gamma}<\nonumber \\
&& <\frac{N^{2-\gamma}c^{1-\gamma}}{2(\gamma-1)}\log(N),
\end{eqnarray}
where we have used $k_M(N)\leq N$.

\section{Derivation of the self-consistent equation}
In this Appendix we derive 
Eqs. (\ref{THEOa}-\ref{THEOe}). Sometimes to indicate
a link we will use the symbol $(i,j)$, or more shortly $ij$.
Let us rewrite explicitly the adimensional Hamiltonian (\ref{H}) as follows
\begin{eqnarray}
\label{H1}
\beta H_{\bm{c}_0,\bm{c}}&=&-\sum_{(i,j)\in\Gamma_0}\left(c_{0;ij}\beta J_{0;ij}+c_{ij}\beta J_{ij}\right)\sigma_{i}\sigma_{j}
\nonumber \\
&& -\sum_{i<j,~(i,j)\notin \Gamma_0}c_{ij}\beta J_{ij}\sigma_{i}\sigma_{j}
-\beta h\sum_i\sigma_i.
\end{eqnarray}
In \cite{MOI} we have introduced the following mapping.
Given a lattice $\mathcal{L}$ with $N=|\mathcal{L}|$ spins, 
and a generic quenched Hamiltonian $H_{\tilde{\bm{J}}}$ 
\begin{eqnarray}
\label{Hd}
\beta H_{\tilde{\bm{J}}}&=&-\sum_{i<j}\beta\tilde{J}_{ij}\sigma_{i}\sigma_{j}-\beta h\sum_i\sigma_i,
\end{eqnarray} 
where the couplings $\{\tilde{J}_{ij}\}$ are distributed according to a given distribution $\{d\tilde{\mu}_{ij}\}$,
let us consider the two following related non random Ising Hamiltonians with labels $\Sigma=$F and $\Sigma=$SG
\begin{eqnarray}
\label{Hd1}
\beta H^{(\Sigma)}&=&-\sum_{(i,j)}\beta\tilde{J}_{ij}^{(\Sigma)}\sigma_{i}\sigma_{j}-\beta h\sum_i\sigma_i,
\end{eqnarray} 
where the effective couplings $\beta\tilde{J}_{ij}^{(\Sigma)}$ are given by
\begin{eqnarray}
\label{Hd2}
\tanh(\beta \tilde{J}_{ij}^{(\Sigma)})=\int d\tilde{\mu}_{ij}(\tilde{J}_{ij})\tanh^{l_\Sigma}(\beta \tilde{J}_{ij}),
\end{eqnarray} 
with $l_\Sigma=1,2$ for $\Sigma=$F or $\Sigma=$SG, respectively.
In \cite{MOI}  we have shown that,  
if the effective couplings $\beta\tilde{J}_{ij}^{(\mathrm{F})}$ or $\beta\tilde{J}_{ij}^{(\mathrm{SG})}$
are at least $\mathop{O}(1/N)$ on the fully connected graph (also called complete graph) $(\mathcal{L},\Gamma_f)$ 
then, in the paramagnetic (P) region, 
the pure model with the effective Hamiltonian $H^{(\Sigma)}$, with $\Sigma=$F or SG, 
gives rise to the same non trivial part of the free energy (see Appendix C) and the same 
correlation functions of the original Hamiltonian $H_{\tilde{\bm{J}}}$,
the stable phase between F and SG being determined by the minimum of the corresponding associated free energies
$f^{(\mathrm{F})}$ or $f^{(\mathrm{SG})}$.  
This in particular gives us the exact critical surfaces paramagnetic-ferro (P-F) 
and paramagnetic spin-glass (P-SG) and, by a simple analytic continuation,
approximations also out of the P region which allow us to get the critical behavior.
The above condition on the effective couplings can be expressed as an infinite dimensionality of the model.
Let us apply the mapping to our small-world scale-free case. The quenched Hamiltonian (\ref{H1})
can be rewritten in the form (\ref{Hd}) where 
\begin{eqnarray}
\label{Hd3}
\tilde{J}_{ij}\defi  
\left\{
\begin{array}{l}
c_{0;ij}J_{0;ij}+c_{ij}J_{ij}, \quad (i,j)\in\Gamma_0,\\
c_{ij}J_{ij}, \quad ~ (i,j)\notin\Gamma_0.
\end{array}
\right.
\end{eqnarray}
By applying Eq. (\ref{Hd2}) to our case with the independent measures $p_0(c_{0;i,j})$ and $p_{ij}(c_{i,j})$
defined by Eqs. (\ref{p0}) and (\ref{hidden}) and with $d\mu_{0ij}=d\mu_0$ and $d\mu_{ij}=d\mu$
being two arbitrary independent measures, we arrive at the following effective couplings
\begin{widetext}
\begin{eqnarray}
\label{Hd4}
\tanh(\beta \tilde{J}_{ij}^{(\Sigma)})=  
\left\{
\begin{array}{l}
(1-p)\int d\mu_0(J_0)\tanh^{l_\Sigma}(\beta J_0)+\mathop{O}(\frac{1}{N}), \quad (i,j)\in\Gamma_0,\\
f(p_i,p_j)\int d\mu(J)\tanh^{l_\Sigma}(\beta J), \quad ~ (i,j)\notin\Gamma_0.
\end{array}
\right.
\end{eqnarray}
\end{widetext}
In particular, for large $N$, in the region $\mathcal{J}$ 
where the factorization (\ref{hidden1}) $f(p_i,p_j)=cNp_ip_j$ takes place,  
from Eqs. (\ref{Hd4}) we get
\begin{eqnarray}
\label{Hd5}
\tilde{J}_{ij}^{(\Sigma)}=  
\left\{
\begin{array}{l}
\tanh^{-1}[t_0^{(\Sigma)}], \quad (i,j)\in\Gamma_0,\\
cNp_ip_jt^{(\Sigma)}, \quad ~ (i,j)\notin\Gamma_0,
\end{array}
\right.
\end{eqnarray}
where $t_0^{(\Sigma)}$ and $t^{(\Sigma)}$ are defined as in Eqs. (\ref{THEOb})-(\ref{THEOe}).
 
We have to evaluate the following partition function
\begin{eqnarray}
\label{ZI0}\nonumber
Z^{(\Sigma)}=\sum_{\{\sigma_i\}}e^{-\beta H_0^{(\Sigma)}+t^{(\Sigma)}\sum_{i<j}f(p_i,p_j)\sigma_{i}\sigma_{j}
+\beta h\sum_i\sigma_i},
\end{eqnarray}
where 
\begin{eqnarray}
\label{H0d}\nonumber
H_0^{(\Sigma)}=-\beta J_0^{(\Sigma)}\sum_{(i,j)\in\Gamma_0}\sigma_{i}\sigma_{j}.
\end{eqnarray}
By using Eq. (\ref{Hd5}) we rewrite $Z^{(\Sigma)}$ as
\begin{eqnarray}
\label{ZI}
Z^{(\Sigma)}&=&
\sum_{\{\sigma_i\}}e^{-\beta H_0^{(\Sigma)}+t^{(\Sigma)}cN\sum_{i<j}p_ip_j\sigma_{i}\sigma_{j}}\nonumber \\
&& \times e^{\beta h\sum_i\sigma_i+\mathop{O}(N^{2-\gamma}\log(N))},
\end{eqnarray}
where $\mathop{O}(N^\alpha\log(N))$ stands for the contributions coming from the
links $(i,j)$ for which the factorization in the second line of Eq. (\ref{Hd5}) is not true
and we have used Eq. (\ref{hidden3c}).
For $N$ large but finite, the corrective term $\mathop{O}(N^\alpha\log(N))$ can be
always neglected, the error per spin being
of order $\mathop{O}(N^{1-\gamma}\log(N))$.

In the following we will suppose that $t^{(\Sigma)}$ is positive.
The derivation for $t^{(\Sigma)}$ negative differs from the other
derivation just for a rotation of $\pi/2$ in the complex $m$-plane, and leads to the
same result one can obtain by analytically continue the equations derived 
for $t^{(\Sigma)}>0$ to the region $t^{(\Sigma)}<0$. 
By using the Gaussian transformation we can rewrite $Z^{(\Sigma)}$ as
\begin{eqnarray}
\label{ZI1}
Z^{(\Sigma)}&=& c_N 
\sum_{\{\sigma_i\}}e^{-\beta H_0^{(\Sigma)}}
\int_{-\infty}^{\infty} d{{m}} ~ e^{-\frac{1}{2} t^{(\Sigma)}c N m^2}
\nonumber \\ 
&& \times 
e^{\sum_i\left(t^{(\Sigma)}c N m p_i + \beta h\right)\sigma_i},
\end{eqnarray}
where $c_N$ is a normalization constant
\begin{eqnarray}\nonumber
c_N = \sqrt{\frac{t^{(\Sigma)}c N}{2\pi}},
\end{eqnarray}
and, in the exponent of Eq. (\ref{ZI1}), 
we have again neglected terms of order $\mathop{O}(1)$.
For finite $N$ we can exchange the integral and the sum over the $\sigma$'s.
By using the definition of the pure model with Hamiltonian $H_0$,
Eq. (\ref{H0}), whose free energy density, for a given coupling $\beta J_0$ and 
for an arbitrary (inhomogeneous) external field $\{\beta h_i\}$,
is indicated by $f_0(\beta J_0,\{\beta h_i\})$, we arrive at
\begin{eqnarray}
\label{ZI2a}
Z^{(\Sigma)}&=& c_N \int_{-\infty}^{\infty} d{{m}} ~ e^{-N L^{(\Sigma)}({{m}})},
\end{eqnarray}
where we have introduced the function 
\begin{eqnarray}
\label{ZI2}
L^{(\Sigma)}({{m}})&=& \frac{1}{2} ct^{(\Sigma)}{{m}}^2 \nonumber \\
&+&\beta f_0 \left(\beta J_0^{(\Sigma)},\{t^{(\Sigma)}c N m p_j + \beta h\}\right).
\end{eqnarray}
By using 
$\partial_{\beta h_i}~N \beta f_0(\beta J_0,\{\beta h_j\})=-m_{0i}(\beta J_0,\{\beta h_j\})$, and
$\partial_{\beta h_j};m_0(\beta J_0,\{\beta h_l\})=\tilde{\chi}_{0;i,j}(\beta J_0,\{\beta h_l\})$,
where $\tilde{\chi}_{0;i,j}\defi \media{\sigma_i\sigma_j}_0-\media{\sigma_i}_0\media{\sigma_j}_0$,
we get  
\begin{eqnarray}
\label{ZI3}\nonumber
&& L^{'(\Sigma)}({{m}})= t^{(\Sigma)}c \left[{{m}} \right. \nonumber \\ && 
\left. -\sum_i m_{0i}\left(\beta J_0^{(\Sigma)},\{t^{(\Sigma)}c N{{m}}p_j + \beta h\}\right)p_i\right], 
\end{eqnarray}
\begin{eqnarray}
\label{ZI4}\nonumber
&& L^{''(\Sigma)}({{m}})= t^{(\Sigma)}c\left[1- t^{(\Sigma)}cN\times \right. \nonumber \\
&& \left. \sum_{i,j}\tilde{\chi}_{0;i,j} \left(\beta J_0^{(\Sigma)},\{t^{(\Sigma)}cN {{m}}p_l + \beta h\}\right)p_i p_j\right]. 
\end{eqnarray}
If the integral in Eq. (\ref{ZI2a}) converges for any $N$,
by performing saddle point integration we see that the saddle point
${{m}}^{(\Sigma)}$ is solution of the equation
\begin{eqnarray}
\label{ZI5}
{{m}}^{(\Sigma)}=\sum_i m_{0i} \left(\beta J_0^{(\Sigma)},
\{t^{(\Sigma)}cN {{m}}^{(\Sigma)}p_j + \beta h\}\right)p_i, 
\end{eqnarray}
so that, if the stability condition
\begin{eqnarray}
\label{ZI6}\nonumber
t^{(\Sigma)}cN
\sum_{i,j}\tilde{\chi}_{0;i,j} \left(\beta J_0^{(\Sigma)},\{t^{(\Sigma)}cN {{m}}^{(\Sigma)}p_l + \beta h\}\right)p_ip_j<1, 
\end{eqnarray}
is satisfied, in the thermodynamic limit we arrive at the following
expression for the free energy density $f^{(\Sigma)}$ of the related Ising model 
\begin{eqnarray}
\label{ZI7}
&&\beta f^{(\Sigma)} = \nonumber \\
&&\left[\frac{t^{(\Sigma)}}{2}c{{m}}^2
+\beta f_0 \left(\beta J_0^{(\Sigma)},\{t^{(\Sigma)}cN {{m}}p_j 
+ \beta h\}\right)\right]_{{{m}}={{m}}^{(\Sigma)}}.
\end{eqnarray}
Similarly, in the thermodynamic limit~\footnote{Note however that finite size effects are responsible
for the critical behavior of the system and, furthermore, as we show in Sec. III D,
the case for $2\mu>1$ must be carefully calculated since the corrections terms responsible for
the critical behavior, for suitable choices of the spin indices $i$ and $j$, may take values up to $\mathop{O}(1)$.},
any correlation function $C^{(\Sigma)}$ of the related Ising model 
is given in terms of the correlation function $C_0$ of the pure
model by the following relation
\begin{eqnarray}
\label{ZI7b}
C^{(\Sigma)} =  C_0\left(\beta J_0^{(\Sigma)},\{t^{(\Sigma)}cN {{m}}p_j 
+ \beta h\}\right)|_{{{m}}={{m}}^{(\Sigma)}}.
\end{eqnarray}

The saddle point solution ${{m}}^{(\mathrm{F})}$ represents the weighted magnetization
(\ref{orderpar}) of the related Ising model, 
as can be checked directly by deriving
Eq. (\ref{ZI7}) with respect to $\beta h$ and by using Eq. (\ref{ZI5}).
For $\Sigma=$SG Eq. (\ref{orderpar}) remains an ansatz. 

If the saddle point equation (\ref{ZI5}) has more stable solutions,
the ``true'' free energy and the ``true'' observable of the related Ising model 
will be given by Eqs. (\ref{ZI7}) and (\ref{ZI7b}), respectively, calculated 
at the saddle point solution which minimizes Eq. (\ref{ZI7}) itself
and that we will indicate with $m^{(\Sigma)}$.

Let us call $\beta_{c0}^{(\Sigma)}$ the inverse critical temperature of the pure
model with coupling $J_0^{(\Sigma)}$ and zero external field, 
possibly with $\beta_{c0}^{(\Sigma)}=\infty$ if no phase transition exists in the pure model.
As stressed in Sec. IIIB, for the pure model 
we use the expression ``critical temperature'' for 
any temperature where the magnetization $m_0$ at
zero external field passes from 0 to a non zero value, continuously or not.
Note that, as a consequence, if $J_0^{(\Sigma)}<0$,
we have formally $\beta_{c0}^{(\Sigma)}=\infty$, 
independently from the fact that
some antiferromagnetic order may be not zero in the pure model.
Let us start to make the obvious observation that a necessary condition for the related Ising model to
have a phase transition at $h=0$ and for a finite temperature, 
is the existence of some paramagnetic region P$^{(\Sigma)}$ where
$m^{(\Sigma)}=0$. By expanding for small $m^{(\Sigma)}=0$ we see from 
the saddle point equation (\ref{ZI5}) that, for $h=0$,
a necessary condition for $m^{(\Sigma)}=0$ to be a solution is that be
$\beta\leq\beta_{c0}^{(\Sigma)}$ for any $\beta$ in P$^{(\Sigma)}$. 
In a few lines we will see however that the inequality must be strict 
if $\beta_{c0}^{(\Sigma)}$ is finite, which in particular excludes
the case $J_0<0$ (for which the inequality to be proved is trivial).
Let us suppose for the moment that be $\beta_c^{(\Sigma)}<\beta_{c0}^{(\Sigma)}$.
For $\beta<\beta_{c0}^{(\Sigma)}$
and $h=0$, the saddle point equation (\ref{ZI5}) has always the trivial 
solution $m^{(\Sigma)}=0$ which, according to the stability condition, is also a stable solution if
\begin{eqnarray}
\label{ZI8}
t^{(\Sigma)}cN
\sum_{i,j}\tilde{\chi}_{0;i,j} \left(\beta J_0^{(\Sigma)},\{0\}\right)p_ip_j<1.
\end{eqnarray}
The solution $m^{(\Sigma)}=0$ starts to be unstable when 
\begin{eqnarray}
\label{ZI9}
t^{(\Sigma)}cN
\sum_{i,j}\tilde{\chi}_{0;i,j} \left(\beta J_0^{(\Sigma)},\{0\}\right)p_ip_j=1.
\end{eqnarray}
Eq. (\ref{ZI9}), together with the constrain $\beta_c^{(\Sigma)}\leq\beta_{c0}^{(\Sigma)}$,
gives the critical temperature of the related Ising model $\beta_c^{(\Sigma)}$.
In the region of temperatures where Eq. (\ref{ZI8}) is violated, 
Eq. (\ref{ZI5}) gives two symmetrical stable 
solutions $\pm m^{(\Sigma)}\neq 0$. 
Furthermore, 
from Eqs. (\ref{ZI5}) and (\ref{ZI9}) we see also that, 
if we make the very plausible assumption that the number of vertices $i$
for which $p_i\geq 1/N$ grows with $N$ as $aN$ (see Eq. (\ref{stat1})), with $a$ asymptotically constant
for $N$ large,
due to the fact that
the pure model has a divergent susceptibility at $\beta_{c0}^{(\Sigma)}$, 
the case $\beta_c^{(\Sigma)}=\beta_{c0}^{(\Sigma)}$
is impossible unless be $t^{(\Sigma)}=0$. 
We have therefore proved that $\beta_c^{(\Sigma)}<\beta_{c0}^{(\Sigma)}$.
Note that for $J_0^{(\Sigma)}\geq 0$ and $\beta<\beta_{c0}^{(\Sigma)}$ 
Eq. (\ref{ZI8}) is violated only for
$\beta>\beta_c^{(\Sigma)}$, whereas for $J_0^{(\Sigma)}<0$ Eq. (\ref{ZI8}) 
in general may be violated also in finite regions of the $\beta$ axis.

\section{Free energy}
Concerning the full expression of the free energy
density, we proceed as follows.
If $\varphi^{(\Sigma)}$ is the high temperature part of the free energy
density $f^{(\Sigma)}$ of the related Ising model that we have solved in Appendix B, then 
\begin{eqnarray}
\label{ZI17} 
&& -\beta f^{(\Sigma)}=  
 \lim_{N\to\infty}\frac{1}{N}\sum_{(i,j)\in\Gamma_0}\log\left[\cosh(\beta J_0^{(\Sigma)})\right]\nonumber \\
&& +\lim_{N\to\infty}\frac{1}{N} \sum_{i<j}\log\left[\cosh\left(ct^{(\Sigma)}Np_ip_j\right)\right] \nonumber \\
&& +\log\left[2\cosh(\beta h)\right] + \varphi^{(\Sigma)}.
\end{eqnarray}
On the other hand, the free energy of the model obeys
\begin{eqnarray}
\label{ZI17v} 
&& -\beta f=\log\left[2\cosh(\beta h)\right] + \varphi\nonumber \\  
&& + \lim_{N\to\infty}\frac{1}{N}\sum_{(i,j)\in\Gamma_0}\int d\mu_0(J_0)(1-p)\log\left[\cosh(\beta J_0)\right]\nonumber \\
&& +\lim_{N\to\infty}\frac{1}{N} \sum_{i<j}\int d\mu(J)\log\left[\cosh\left(\beta J\right)\right]cNp_ip_j.
\end{eqnarray}
Therefore, by using the mapping $\varphi=\varphi^{(\Sigma)}/l_{\Sigma}$, 
and $\beta f^{(\Sigma)}=L^{(\Sigma)}(m^{(\Sigma)})$, where $L^{(\Sigma)}$
is given by Eq. (\ref{ZI2}), and $m^{(\Sigma)}$ is the solution of the self-consistent Eq. (\ref{THEOa}),
and by choosing $\Sigma$ according to which is minimum between $L^{(\mathrm{F})}(m^{(\mathrm{F})})$ 
and $L^{(\mathrm{SG})}(m^{(\mathrm{SG})})$, 
comparing Eq. (\ref{ZI17}) with Eq. (\ref{ZI17v}) we get 
the total free energy $\beta f$. It is clear however (as already anticipated) 
that the only part of the free energy which depends on the order parameter and that
is therefore responsible for the critical behavior of the system and the correlation functions is $\varphi$.
The rest of the free energy is important only to calculate the total specific heat.
\section{Proof of Eq. (\ref{THEOp8g})}
Let us start to express the partition function of the pure
model in the high temperature expansion. 
In general, for the partition function $Z$ of an Ising model having a set of links $b\in\Gamma$
taking the couplings $\{J_b\}$ and in the presence of arbitrary external fields $\{h_i\}$, we have 
\begin{eqnarray}
\label{Zh}
Z\left(\{J_b\};\{h_i\}\right) &=&
\prod_{b\in\Gamma} \cosh\left(\beta J_b\right) \prod_{i=1}^N \cosh\left(\beta h_i\right)\nonumber \\ 
&& \times \sum_{\{\sigma_i\}}
\prod_{b\in\Gamma} \left[1+\sigma_{i_b}\sigma_{j_b}\tanh\left(\beta J_b\right)\right]\nonumber \\
&& \times \prod_{i=1}^N \left[1+\sigma_i\tanh\left(\beta h_i\right)\right],
\end{eqnarray} 
where $i_b$ and $j_b$ are the two sites linked by the link $b$.
It is not difficult to recognize that $Z$ can be rewritten as a sum over paths
as follows
\begin{eqnarray}
\label{Zhb}
Z\left(\{J_b\};\{h_i\}\right) &=& 
\prod_{b\in\Gamma} \cosh\left(\beta J_b\right) \prod_{i=1}^N \cosh\left(\beta h_i\right)\nonumber \\ 
&& \times \sum_{\gamma\in\mathcal{T}}\prod_{b\in\gamma} t_b\prod_{i\in\partial \gamma} t_i, 
\end{eqnarray} 
where: $\mathcal{T}$ is the set of all possible multi-paths on $\Gamma$,
including then all the possible combinations of closed and open paths; $\partial \gamma$ stands
for the subset of vertices which belong to the border of the multipath $\gamma$ 
(if it has at least one open path component);
and we have introduced the short notations $t_b=\tanh(\beta J_b)$ and $t_i=\tanh(\beta h_i)$. 
Note that the cardinality $|\partial \gamma|$ is always an even number.
We want now to calculate the average magnetization $\media{\sigma_i}$. 
By derivating Eq. (\ref{Zhb}) with respect to $\beta h_i$ we get
\begin{eqnarray}
\label{Zhc}
\media{\sigma_i} &=& t_i+(1-t_i^2)
\frac{\sum_{\gamma\in\mathcal{T}^{(i)}}\prod_{b\in\gamma} t_b\prod_{j\in\partial \gamma\setminus i} t_j}
{\sum_{\gamma\in\mathcal{T}}\prod_{b\in\gamma} t_b\prod_{j\in\partial \gamma} t_j},
\end{eqnarray} 
where $\mathcal{T}^{(i)}$ stands for the subset of $\mathcal{T}$ having at least one
open path component which passes through the vertex $i$. 
As done in Sec. IIID, we cannot expand the terms $t_i$ for small $h_i$, but 
we can neglect $\mathop{(t_i)^2}$ terms. In other words, 
we expand $\media{\sigma_i}$ at the least non zero 
order - not in the $\{h_j\}$, and nor in the $\{t_b\}$ -
but in the $\{t_j\}$.
Within this approximation Eq. (\ref{Zhc}) becomes
\begin{eqnarray}
\label{Zhd}
\media{\sigma_i} &=& t_i + 
\frac{\sum_{\gamma\in\mathcal{T}_{1}^{(i)}}\prod_{b\in\gamma} t_b t_{j^{(i)}}}
{\sum_{\gamma\in\mathcal{C}}\prod_{b\in\gamma} t_b}
+\mathop{O}(\{t_j^2\}), 
\end{eqnarray} 
where now $\mathcal{T}_{1}^{(i)}$
stands for the subset of $\mathcal{T}$ having one and only one
open path component which passes through the vertex $i$, 
$j^{(i)}$ is second end of this path component passing through $i$, 
and $\mathcal{C}$ stands for the set of all the closed multi-paths on $\Gamma$.
Notice that, as anticipated, this latter definition makes the calculation exact with respect 
to the presence of loops of any length and taking any coupling. 
Similarly, for the connected correlation function, by deriving once more with respect to $\beta h_j$ we get
\begin{eqnarray}
\label{Zhe}
\tilde{\chi}_{ij} &=& \delta_{i,j} + 
\frac{\sum_{\gamma\in\mathcal{T}_{1}^{(i,j)}}\prod_{b\in\gamma} t_b}
{\sum_{\gamma\in\mathcal{C}}\prod_{b\in\gamma} t_b}
+\mathop{O}(\{t_j^2\}), 
\end{eqnarray} 
where $\mathcal{T}_{1}^{(i,j)}$
stands for the subset of $\mathcal{T}$ having one and only one
open path component which passes through both the vertices $i$ and $j$.
From Eq. (\ref{Zhe}) one can obtain the susceptibility $\tilde{\chi}$
up to $\mathop{O}(\{t_j^2\})$ terms by summing over $i$ and $j$ and dividing by $N$.
In particular, for a regular lattice we have
\begin{eqnarray}
\label{Zhe}
\tilde{\chi} &=& 1 + 
\frac{\sum_{\gamma\in\mathcal{T}_{1}^{(i_0)}}\prod_{b\in\gamma} t_b}
{\sum_{\gamma\in\mathcal{C}}\prod_{b\in\gamma} t_b}
+\mathop{O}(\{t_j^2\}), 
\end{eqnarray} 
where $i_0$ is an arbitrary vertex chosen as reference.
If we now, for a regular lattice, plug in Eq. (\ref{Zhd}) in the self-consistent equation (\ref{THEOa}), 
use the definition (\ref{THEOp4b}) and the property (\ref{THEOp5}), 
we get the bound 
\begin{eqnarray}
\label{Zhf}
m &<& g(m)+ 
\frac{\sum_{\gamma\in\mathcal{T}_{1}^{(i_0)}}\prod_{b\in\gamma} t_b g(m)}
{\sum_{\gamma\in\mathcal{C}}\prod_{b\in\gamma} t_b}
+\mathop{O}(\{t_j^2\}), 
\end{eqnarray} 
which, by using (\ref{Zhe}), leads immediately to Eq. (\ref{THEOp8g}).
\section{Derivation of Eq. (\ref{THEOC2b})}
Eqs. (\ref{ZI2a})-(\ref{ZI5}) are already in a form able to take into
account the presence of an arbitrary inhomogeneous external field, $\{h_j\}$;
in these equations we have simply to substitute everywhere in their arguments $\{h\}$ with $\{h_j\}$.
Then, by deriving $L^{(\Sigma)}({{m}})$ with respect to $\beta h_i$ and by using, as in Appendix B, 
$\partial_{\beta h_i}~N \beta f_0(\beta J_0,\{\beta h_j\})=-m_{0i}(\beta J_0,\{\beta h_j\})$, and
$\partial_{\beta h_j};m_0(\beta J_0,\{\beta h_l\})=\tilde{\chi}_{0;i,j}(\beta J_0,\{\beta h_l\})$,
and the self-consistent equation for the order parameter ${{m}}^{(\Sigma)}$,
we get immediately
\begin{eqnarray}
\label{ZI5C}
{{m}}^{(\Sigma)}_i=m_{0i} \left(\beta J_0^{(\Sigma)},\{t^{(\Sigma)}cN {{m}}^{(\Sigma)}p_j + \beta h_j\}\right), 
\end{eqnarray}
which confirms Eq. (\ref{THEOC2b}) for the correlation functions of order $k=1$.
Then, by deriving in turn ${{m}}^{(\Sigma)}_i$ with respect to $\beta h_j$, and by using
\begin{widetext}
\begin{eqnarray}
\label{THEOC2bC}
\frac{\partial m^{(\Sigma)}}{\partial (\beta h_j)}&=&  
\frac{\sum_np_n\tilde{\chi}_{0;n,j} (\beta J_0^{(\Sigma)};\{Np_qct^{(\Sigma)}m^{(\Sigma)}+\beta h\})}
{1-ct^{(\Sigma)}N\sum_{l,n}\tilde{\chi}_{0;l,n}
\left(\beta J_0^{(\Sigma)};\{Np_qct^{(\Sigma)}m^{(\Sigma)}+\beta h\}\right)p_lp_n},
\end{eqnarray}
\end{widetext}
we reach immediately Eq. (\ref{THEOC2b}).



\end{document}